\documentclass[floats,floatfix,showpacs,amssymb,prd,twocolumn,superscriptaddress,nofootinbib]{revtex4-1}

\usepackage{graphicx,epsf, epsfig, amssymb}% Include figure files
\usepackage{bm}
\usepackage{longtable}
\usepackage{color}
\usepackage[breaklinks]{hyperref}
\usepackage{amsfonts,amsmath,amssymb,mathrsfs}

\def\be{\begin{equation}}
\def\ee{\end{equation}}
\def\beq{\begin{eqnarray}}
\def\eeq{\end{eqnarray}}
\def\f{\frac}

\newcommand{\nn}{\nonumber}
\def\ii{{\rm i}}
\def\tr{{\hat r}}

\newcommand{\tA}{\tilde{A}}

\newcommand{\bgaln}{\begin{align}}
\newcommand{\bgeq}{\begin{equation}}

\newcommand{\Rmnum}[1]{\expandafter\@slowromancap\romannumeral #1@}

\begin{document}

\title{Numerical simulations of single and binary black holes in scalar-tensor theories:\\
circumventing the no-hair theorem}

%%%%%%%%%%%%%%%%%%%%%%%%%%%%%%%%%%%%%%%%%%%%%%%%%%%%%%%%%%%%%%%%%%%%%%%%%%%%%%%%
%%%%%%%%%%%%%%%%%%%%  AUTHORS  %%%%%%%%%%%%%%%%%%%%%%%%%%%%%%%%%%%%%%%%%%%%%%%%%
%%%%%%%%%%%%%%%%%%%%%%%%%%%%%%%%%%%%%%%%%%%%%%%%%%%%%%%%%%%%%%%%%%%%%%%%%%%%%%%%
\author{Emanuele Berti} \email{berti@phy.olemiss.edu}
\affiliation{Department of Physics and Astronomy, The University of
  Mississippi, University, MS 38677, USA}
\affiliation{California Institute of Technology, Pasadena, CA 91109, USA}

\author{Vitor Cardoso} \email{vitor.cardoso@ist.utl.pt}
\affiliation{Department of Physics and Astronomy, The University of
Mississippi, University, MS 38677, USA}
\affiliation{CENTRA, Departamento de F\'{\i}sica, Instituto Superior
T\'ecnico, Universidade T\'ecnica de Lisboa - UTL, Av.~Rovisco Pais
1, 1049 Lisboa, Portugal}
\affiliation{Perimeter Institute for Theoretical Physics
Waterloo, Ontario N2J 2W9, Canada}
\affiliation{Faculdade de F\'{\i}sica, Universidade 
Federal do Par\'a, 66075-110, Bel\'em, Par\'a, Brazil}

\author{Leonardo Gualtieri} \email{Leonardo.Gualtieri@roma1.infn.it}
\affiliation{Dipartimento di Fisica, Universit\`a di Roma ``Sapienza''
  \& Sezione, INFN Roma1, P.A. Moro 5, 00185, Roma, Italy}

\author{Michael Horbatsch} \email{mhorbats@olemiss.edu}
\affiliation{Department of Physics and Astronomy, The University of
  Mississippi, University, MS 38677, USA}

\author{Ulrich Sperhake} \email{sperhake@tapir.caltech.edu}
\affiliation{Department of Applied Mathematics and Theoretical Physics, Centre for Mathematical Sciences, University of Cambridge, Wilberforce Road, Cambridge CB3 0WA, UK}
\affiliation{Institut de Ci{\'e}ncies de l'Espai (CSIC-IEEC), Facultat
  die Ci{\'e}ncies, Campus UAB, E-08193 Bellaterra, Spain}
\affiliation{Department of Physics and Astronomy, The University of
  Mississippi, University, MS 38677, USA}
\affiliation{California Institute of Technology, Pasadena, CA 91109, USA}
\affiliation{CENTRA, Departamento de F\'{\i}sica, Instituto Superior
  T\'ecnico, Universidade T\'ecnica de Lisboa - UTL, Av.~Rovisco Pais
  1, 1049 Lisboa, Portugal}

\date{\today}

\begin{abstract}
Scalar-tensor theories are a compelling alternative to general
relativity and one of the most accepted extensions of Einstein's
theory. Black holes in these theories have no hair, but could grow
``wigs'' supported by time-dependent boundary conditions or spatial
gradients.  Time-dependent or spatially varying fields lead in general
to nontrivial black hole dynamics, with potentially interesting
experimental consequences. We carry out a numerical investigation of
the dynamics of single and binary black holes in the presence of
scalar fields. In particular we study gravitational and scalar
radiation from black-hole binaries in a constant scalar-field
gradient, and we compare our numerical findings to analytical models.
In the single black hole case we find that, after a short transient,
the scalar field relaxes to static configurations, in agreement with
perturbative calculations. Furthermore we predict analytically (and
verify numerically) that accelerated black holes in a scalar-field
gradient emit scalar radiation. For a quasicircular black-hole binary,
our analytical and numerical calculations show that the dominant
component of the scalar radiation is emitted at twice the binary's
orbital frequency.
\end{abstract}
\maketitle
%%%%%%%%%%%%%%%%%%%%%%%%%%%%%%%%%%%%%%%%%%%%%%%%%%%%%%%%%%%%%
%%%%%%%%%%%%%%%%%%%%%%%%%%%%%%%%%%%%%%%%%%%%%%%%%%%%%%%%%%%%%
\section{Introduction}
%%%%%%%%%%%%%%%%%%%%%%%%%%%%%%%%%%%%%%%%%%%%%%%%%%%%%%%%%%%%%
%%%%%%%%%%%%%%%%%%%%%%%%%%%%%%%%%%%%%%%%%%%%%%%%%%%%%%%%%%%%%

Scalar fields are ubiquitous in physics, either as a proxy for more
complex interactions or as fundamental quantities in their own right.
For example, one of the best studied modifications to general
relativity is scalar-tensor gravity, in which space-time curvature
couples to scalar fields (which are sufficiently light 
to be relevant for astrophysics and/or cosmology).
The historical development of this theory goes back to the 1940s, and
involves several research groups with different views on the physical
interpretation of the scalar degree of freedom
\cite{Goenner:2012cq,Brans:2005ra,Singh:1983qp}.  In recent times,
interest in scalar-tensor gravity has been driven by theoretical
attempts to unify gravity with quantum mechanics at high energies and
solve the cosmological constant and hierarchy problems, as well as
observations of the cosmic microwave background and the
highly-anticipated direct detection of gravitational waves
\cite{Fujii:2003pa,faraonicosmbook,sctenscosmbook}.

The simplest version of scalar-tensor gravity is Brans-Dicke theory
\cite{Brans:1961sx,Dicke:1961gz,Brans:1962zz}, which consists of a
single massless scalar $\phi$, whose coupling to curvature is
controlled by a dimensionless parameter $\omega_{\rm
  BD}$. Generalizations include varying scalar-curvature couplings
$\omega(\phi)$ and a scalar potential $V(\phi)$ (``Bergmann-Wagoner''
theories \cite{Bergmann:1968ve,Wagoner:1970vr}) 
as well as the possibility of multiple interacting scalar fields: see
e.g.  \cite{Fujii:2003pa,Damour:1992we,Chiba:1997ms} for comprehensive
treatments of the subject.
These generalizations (commonly referred to as ``scalar-tensor
theories'') are compelling due to their simplicity, but (perhaps as a
consequence) they are also very well constrained observationally. The
classic book by Will \cite{Will:1993ns} presents an overview of the
subject, and comprehensive reviews of the state of the art in
experimental tests of gravitational theories can be found in
\cite{Will:2005va,Clifton:2011jh}.

In this paper we shall focus, for simplicity, on scalar-tensor
theories involving a single scalar field. In general relativity,
because of the conservation of total momentum for isolated systems,
gravitational radiation is quadrupolar in nature. A possible smoking
gun of scalar-tensor gravity is the existence of dipole radiation,
essentially due to violations of the equivalence principle. Solar
System experiments and observations of binary pulsar systems place
strong constraints on the coupling functions of the theory
\cite{Will:1989sk,Damour:1992we,Will:1993ns,Damour:1993hw,Damour:1995kt,Damour:1996ke,Damour:1998jk,Will:2005va,Damour:2007uf,Bhat:2008ck,Lazaridis:2009kq,EspositoFarese:2009ta,Alsing:2011er,Freire:2012mg},
and other interesting constraints may come from the direct observation
of gravitational radiation from binary systems in the near future
\cite{Will:1994fb,Damour:1998jk,Scharre:2001hn,Will:2004xi,Berti:2004bd,Berti:2005qd,Yagi:2009zm,Yunes:2011aa,Berti:2012bp}.
Despite all of these observational constraints, striking and
potentially observable astrophysical phenomena are still possible in
these theories. Such phenomena include superradiant instabilities (see
e.g. \cite{Arvanitaki:2010sy,Dubovsky:2010je,Yoshino:2012kn,Witek:2012tr}
for discussions of this phenomenon in the context of the ``string
axiverse'' scenario \cite{Arvanitaki:2009fg}) and the related
possibility of floating orbits around rotating black holes (BHs)
\cite{Cardoso:2011xi}.

%%%%%%%%%%%%%%%%%%%%%%%%%%%%%%%%%%%%%%%%%%%%%%%%%%%%%%
\subsection{Classical no-hair theorems}
%%%%%%%%%%%%%%%%%%%%%%%%%%%%%%%%%%%%%%%%%%%%%%%%%%%%%%

Theoretical studies impose remarkable constraints and limitations on
scalar-tensor theories. First of all, the famous BH no-scalar-hair
theorems first proved by Hawking \cite{Hawking:1972qk}, Thorne and
Dykla \cite{1971ApJ...166L..35T}, and Chase \cite{Chase:1970} state
that stationary BH solutions in Brans-Dicke theory are the same as
those in general relativity. These results have been generalized and
expanded upon by many authors. For example, an extension to multiple
scalars has been established by Heusler \cite{Heusler:1995qj}, and an
extension to Bergmann-Wagoner and $f(R)$ theories has been established
by Sotiriou and Faraoni \cite{Sotiriou:2011dz}. The no-scalar-hair
theorems have also been confirmed by numerical studies of
gravitational collapse
\cite{Scheel:1994yr,Scheel:1994yn,Shibata:1994qd,Harada:1996wt,Novak:1997hw,Kerimo:1998qu}.
More generally, it has been observed that the Kerr metric is a
solution in a wide class of gravity theories
\cite{Psaltis:2007cw}. However, the fact that stationary vacuum
solutions of scalar-tensor theories agree with those of general
relativity does not mean that the dynamics of BHs in these theories
must be the same \cite{Scheel:1994yn,Kerimo:2000gc,Barausse:2008xv,Berti:2013uda}.

For a comprehensive discussion and literature survey of no-hair
theorems, the reader is referred to the reviews of Bekenstein
\cite{Bekenstein:1996pn} and Chru\'{s}ciel, Costa, and Heusler
\cite{Chrusciel:2012jk}. Although the literature is vast, there are
two basic assumptions lying at the heart of most no-hair theorems.
The first is that of stationarity, whose necessity has been
demonstrated by Jacobson's ``Miracle Hair Growth Formula,'' a
perturbative construction of a hairy BH with time-dependent scalar
boundary conditions \cite{Jacobson:1999vr}.

The second assumption is the truncation of the scalar-tensor action to
second order in the derivative expansion.  At this level, the most general
action [Eq.~(\ref{eq:sctens_action_jf}) in the single-scalar case] is very
simple and contains only three terms, whereas scalar-tensor gravity at
the four-derivative level is too complicated to be studied in complete
generality, and thus, attention is often restricted to particular models.

One such model is quadratically modified gravity, whose action
contains all possible terms quadratic in the Riemann tensor, coupled
to a scalar. In this theory, BHs have been studied perturbatively, and
solutions with scalar hair have been found
\cite{Yagi:2012ya,Yunes:2009hc,Konno:2009kg,Yunes:2011we,Pani:2011gy}. Moreover,
in the special case of a scalar coupled to a topological invariant --
namely, Einstein-dilaton-Gauss-Bonnet (EDGB) or Dynamical Chern-Simons
(DCS) gravity -- a no-hair theorem for neutron stars has been
established \cite{Yagi:2011xp}. The conclusion here is that
spherically-symmetric neutron stars have vanishing scalar monopole
moment, but higher-order scalar multipoles need not vanish.  However,
the presence of derivatives higher than second order in the field
equations severely complicates the implementation of numerical
simulations.

Although four-derivative actions generically lead to field equations
with more than two derivatives, there are some noteworthy exceptions.
One is the Einstein-Skyrme system, a nonlinear sigma model with
target-space $SU(2)$, containing a term in the action quartic in
scalar derivatives, and admitting linearly-stable BH solutions with
scalar hair which have been described numerically in
\cite{Droz:1991cx,Heusler:1991xx,Heusler:1992av,Shiiki:2005pb}.
Another model with second-order field equations is the galileon
\cite{galileon}, which is related to both higher-dimensional Lovelock
gravity \cite{VanAcoleyen:2011mj} and massive gravity
\cite{Luty:2003vm,deRham:2010kj}. 
It satisfies Solar System constraints by means of the Vainshtein
mechanism \cite{vainshtein}, and a no-galileon-hair theorem for
spherically-symmetric BHs has been recently established
\cite{Hui:2012qt}. Another interesting example is Bergmann-Wagoner
scalar-tensor gravity coupled to non-linear electrodynamics, where the
non-vanishing trace of the electromagnetic stress-energy tensor enters
as a source into the scalar-field equation. This allows for stable
asymptotically-flat BHs with scalar hair, which have been studied
numerically in
\cite{Stefanov:2007qw,Stefanov:2007eq,Stefanov:2007bn,Stefanov:2009qd,Stefanov:2009zza,Doneva:2010ke}.

%%%%%%%%%%%%%%%%%%%%%%%%%%%%%%%%%%%%%%%%%%%%%%%%%%%%%%
\subsection{A generalized no-hair theorem}
%%%%%%%%%%%%%%%%%%%%%%%%%%%%%%%%%%%%%%%%%%%%%%%%%%%%%%

The ``classical'' no-hair theorems described in the preceeding
paragraphs, which are statements about stationary vacuum space-times,
have been extended to the context of a BH binary system.  Employing
the ``generalized EIH'' formalism developed by Eardley
\cite{Eardley:1975}, Will and Zaglauer \cite{Will:1989sk} have shown
that the leading-order post-Newtonian (PN) dynamics of a BH binary in
Brans-Dicke theory is indistinguishable from that in general
relativity.  Recently, this result has been extended to general
scalar-tensor theories
in the extreme mass-ratio limit \cite{Yunes:2011aa} 
and it has been shown to hold up to 2.5PN order for generic mass ratio
\cite{Mirshekari:2013vb}.  Thus, even a dynamical (vacuum) spacetime
with two interacting BHs does not have scalar hair in the PN limit. We
can regard this conclusion as a ``generalized no-hair theorem.''

The generalized no-hair theorem relies on the assumption that the
binary system is isolated, in the sense that cosmological and
environmental effects (say, due to the galactic background surrounding
the BHs) are neglected.  More precisely, it is assumed that: (1) there
is no matter in the system, (2) the scalar field has zero potential,
(3) the scalar-tensor action is truncated to second order in the
derivative expansion, and (4) the metric is asymptotically flat (in
all conformal frames) and the scalar is asymptotically constant.

The vacuum assumption (1) can be relaxed either by considering BHs in
astrophysical environments or by considering compact stars, which are
affected by the well-known spontaneous scalarization phenomenon
(cf.~\cite{Barausse:2012da} for recent numerical studies).
Other recent numerical studies created a scalar-field ``bubble''
around the binary by using a nonvanishing potential, i.e. relaxing
assumption (2). They found that the scalar-field bubble is rapidly
accreted by the BHs, modifying the binary dynamics
\cite{Healy:2011ef}.
As for assumption (3), compact binary dynamics in quadratically
modified gravity has been studied analytically in a perturbative
framework, where one takes the point of view that the model should be
considered as an effective low-energy theory
\cite{Yagi:2011xp,Yagi:2012vf,Yagi:2013mbt}. Whether these theories
are well posed for numerical evolutions is currently a matter of
debate.  We will not consider this problem in the present paper, but
it is an interesting topic for future research.

Relaxing either assumption (3) or assumption (4) introduces a new
length (or time) scale in the BH binary dynamics. In the case of
assumption (3) this scale is determined by the Compton wavelength of a
heavy particle, whose square enters into coefficients of
four-derivative terms in the action.  In the case of assumption (4),
the new scale is determined by cosmological and/or galactic effects.
A priori, it is not obvious which of these effects is dominant, and
thus it is worthwhile to explore both possibilities. To our knowledge,
the relaxation of assumption (4) has not been explored in the
literature, and it is the main focus of our paper.

Asymptotic flatness of the metric is only an approximation to the
dynamics of an astrophysical binary. Observations show that the
Universe is expanding on timescales which are very large, but
nevertheless finite with respect to astrophysical BH binary
evolution. As shown in \cite{Jacobson:1999vr,Horbatsch:2011ye},
imposing time-varying boundary conditions endows the BHs in a binary
with scalar charge, and therefore the binary can emit dipole scalar
radiation. Furthermore, many cosmological models consider the
existence of background scalars which can be anchored on matter
\cite{PhysRevD.50.3650,Hu:2000ke,Matos:1998vk,Schunck:1998nq,Sahni:1999qe,Matos:2000ki,Arbey:2001qi,Arbey:2003sj,Boehmer:2007um,Alcubierre:2001ea}. In
this case, one can for instance conceive of a BH binary evolving in
the background of a nearly-static but nonuniform scalar field anchored
on the galactic matter.
The characteristic lengthscale of such a scalar-field profile would be
much larger than the binary separation, and therefore it would have
the same effect on the dynamical evolution of the binary system as the
enforcement of boundary conditions which are not asymptotically flat.
Finally, BH dynamics in the background of scalar fields could also be
relevant to understanding accretion inside hypothetical supermassive
boson stars, where huge scalar-field gradients are expected
\cite{Macedo:2013qea}.

Scalar-field gradients can therefore allow us to circumvent the
generalized no-hair theorem, i.e., to have a spacetime which contains
only BHs and still emits scalar radiation.  Indeed, as we shall
discuss below [cf. Eq.~(\ref{ScalarCharge}), Section \ref{appsol}], in
the presence of a spatially-varying
scalar-field profile $\varphi(\vec{x})$, a nonrotating BH of mass $M$
with world-line $(t,\vec{x}(t))$ would have a scalar
charge\footnote{The scalar charge $Q$ and mass $M$ are Einstein-frame
  quantities, and geometrical units $G=c=1$ are employed, where $G$ is
  the Einstein-frame bare gravitational constant. The result quoted
  here also assumes that the BH motion relative to the scalar-field
  profile is ``slow'', in the sense that $M(d\vec{x}/dt) \cdot
  \vec{\nabla} \varphi \ll 1$. }
\beq
\label{Qv}
Q(t)&=& 
4M^{2} \, \frac{d \vec{x}(t)}{dt} \cdot \vec{\nabla} \varphi (\vec{x}(t)) 
\\
&=&8 \pi \sigma M^{2} \, \frac{d \vec{x}(t)}{dt} \cdot \hat{z} 
\,,\nn
\eeq
where in the second line we have assumed that the scalar-field
gradient is directed along the $z$-axis, and we have parametrized its
magnitude by a real parameter $\sigma$. If the BH is accelerated, or
if the scalar gradient is nonuniform, the scalar charge would evolve
in time, yielding scalar radiation.
As we show in Appendix \ref{cosmo}, for a stellar-mass BH
($M=10\,M_\odot$) moving near the galactic center a typical
scalar-field gradient is $M\sigma\sim 10^{-15}$. For a supermassive BH
with $M=10^9\,M_\odot$ a typical gradient could be as large as
$M\sigma\sim 10^{-7}$, comparable in order of magnitude to the
numerical simulations presented in this paper.

%%%%%%%%%%%%%%%%%%%%%%%%%%%%%%%%%%%%%%%%%%%%%%%%%%%%%%
\subsection{Executive summary and plan of the paper}
%%%%%%%%%%%%%%%%%%%%%%%%%%%%%%%%%%%%%%%%%%%%%%%%%%%%%%

The main goal of this work is to explore the consequences of the
presence of a scalar-field gradient, which is equivalent to imposing
nontrivial boundary conditions on the dynamics of a BH binary, and to
verify numerically whether, as suggested by Eq.~(\ref{Qv}), such a
setup can allow scalar radiation from a BH binary system in
scalar-tensor theory. Here we present an executive summary of our main
results and an outline of the paper.

In Section~\ref{sec:framework} we lay out our theoretical framework by
introducing generic scalar-tensor theories and presenting the
relations that allow us to transform between the Jordan frame (where
physical quantities should be computed) and the Einstein frame (where
we will perform our calculations). In particular, we show how
gravitational radiation in the Jordan frame can be computed from a
knowledge of the Newman-Penrose scalars in the Einstein frame.  

In Section~\ref{sec:analytic} we introduce analytical approximations
for scalar fields in the background of single and binary BH
spacetimes.  These approximations are useful to validate (and provide
insight into) our numerical simulations. In fact, numerical evolutions
of initial data corresponding to a single black hole in a scalar
gradient show that, after a short transient, the scalar field relaxes
to the static configurations predicted by these perturbative
calculations. For a quasicircular black-hole binary, in
Section~\ref{sec:analytic} we show analytically that the dipole
component of the scalar radiation is emitted at twice the binary's
orbital frequency. This prediction is validated by our numerical
simulations, which also show that the dipole component dominates the 
scalar emission.

In Section \ref{sec:numerical} we present the details of our numerical
implementation. The results of our simulations are discussed and
compared with analytical results in Section \ref{sec:results}. In
Section \ref{sec:conclusions} we summarize our findings and point out
possible directions for future work.

To improve readability, in the Appendices we collect technical
material that illustrates various important points of our
analysis. Appendix \ref{app:boosted} shows that a BH moving with
constant velocity in a uniform scalar-field gradient does not emit
scalar radiation. Appendix \ref{app:binaryscaleq} (which is
complementary to Section \ref{appsol}) collects some lengthy formulas
illustrating the structure of gravitational radiation from a BH binary
moving in a scalar-field gradient. In Appendix
\ref{sec:evolution_equations} we provide explicit expressions for the
evolution equations used in our numerical code. In Appendix
\ref{cosmo} we estimate the order of magnitude of the scalar-field
gradients expected in scalar-field models of dark matter.

%%%%%%%%%%%%%%%%%%%%%%%%%%%%%%%%%%%%%%%%%%%%%%%%%%%%%%%%%%%%%
\section{Theoretical Framework}
\label{sec:framework}
%%%%%%%%%%%%%%%%%%%%%%%%%%%%%%%%%%%%%%%%%%%%%%%%%%%%%%%%%%%%%

We focus on general single-scalar-tensor theories in vacuum with
vanishing scalar potential, and at most two derivatives in the
action. These theories are equivalent to Einstein's theory extended to
include a minimally coupled scalar field with vanishing potential.
This statement (which will be clarified below) has a nontrivial
consequence: the addition of minimally coupled scalars to Einstein's
gravity allows one to study a multitude of scalar-tensor theories at
once.  For this reason our simple framework offers an opportunity to
take a glimpse at a rather broad spectrum of physics beyond Einstein's
theory. 

Our starting point is the action of a general
scalar-tensor theory for a single scalar field $\phi$, written as
\be
\label{eq:sctens_action_jf}
S=\int d^4x\frac{\sqrt{-g}}{16\pi G} \left(F(\phi)R-8\pi G Z(\phi)g^{\mu\nu}\partial_{\mu}\phi\partial_{\nu}\phi-U(\phi)\right)\,,
\ee
where $R$ is the Ricci scalar associated to the metric $g_{\mu\nu}$,
and $F(\phi), Z(\phi)$ and $U(\phi)$ are arbitrary functions (see
e.g.~\cite{Fujii:2003pa} and references therein). This form of the
action corresponds to the choice of the so-called ``Jordan frame'',
where the scalar field is nonminimally coupled with gravity and all
other matter fields obey the equivalence principle. The dynamics of
matter fields would be described by an additional term $S_{\rm matt}$
on the right-hand side, that we set equal to zero because we are
interested in BHs in vacuum.

Our numerical evolutions are more easily performed in the so-called
Einstein-frame representation, that is related to the Jordan-frame
representation by a conformal rescaling of the metric. In the Einstein
frame, the scalar field is minimally coupled with the metric tensor
and it affects the scalar-matter coupling in the matter action $S_{\rm
  matt}$. Working in the Einstein frame is convenient because we focus
on pure BH spacetimes, i.e., we set $S_{\rm matt}=0$. 
Since we are interested in the effect of boundary conditions, we shall
assume for simplicity (as in \cite{Horbatsch:2011ye}, and at variance
with \cite{Healy:2011ef}) that the effect of the scalar-field
potential is negligible: $U(\phi)=0$.

%%%%%%%%%%%%%%%%%%%%%%%%%%%%%%%%%%%%%%%%%%%%%%%%%%%%%%%%%%%%%
\subsection{From Jordan to Einstein and back}
%%%%%%%%%%%%%%%%%%%%%%%%%%%%%%%%%%%%%%%%%%%%%%%%%%%%%%%%%%%%%
With $U(\phi)=0$, the explicit transformations that recast the
previous action in the Einstein frame are~\cite{Damour:1996ke}
\begin{eqnarray}
g^E_{\mu\nu}&=&F(\phi)g_{\mu\nu}\,,
\label{JEtransf1}
\\
\varphi(\phi)&=&\int d\phi\,\left[\frac{3}{2}\frac{F'(\phi)^2}{F(\phi)^2}+\frac{8\pi G Z(\phi)}{F(\phi)}\right]^{1/2}\,,
\label{JEtransf2}
\\
A(\varphi)&=&F^{-1/2}(\phi)\,.
\label{JEtransf3}
\end{eqnarray}
The Einstein-frame action is then
\begin{equation}
S=\frac{1}{16\pi G}\int\left[R^{E}-g_{E}^{\mu \nu}\partial_{\mu}\varphi \partial_{\nu} \varphi\right]\sqrt{-g^{E}}d^4x\,,
\label{action}
\end{equation}
and it leads to the following equations of motion:
\beq
\label{Eeq}
G_{\mu\nu}^{E}&=&\partial_{\mu} \varphi \partial_{\nu}\varphi
                 -\frac{1}{2}g_{\mu\nu}^{E}
g_{E}^{\rho \sigma} \partial_\rho \varphi \partial_\sigma \varphi\,,\\
\label{Seq}
\Box^{E}\varphi&=&0\,,
\eeq
where the label $E$ denotes quantities built out of the Einstein-frame
metric $g_{\mu \nu}^{E}$.  This will be the starting point of our
analysis. It is important to stress that, even though we are formally
investigating a minimally coupled theory, our results are in principle
applicable to a wide range of scalar-tensor theories. By working in
the Einstein frame we can focus on quantities that depend on the
intrinsic properties of the binary, rather than quantities that would
be measured by gravitational-wave detectors. The latter may be
obtained for any specific theory using the transformation from the
Einstein frame to the Jordan frame.

%%%%%%%%%%%%%%%%%%%%%%%%%%%%%%%%%%%%%%%%%%%%%%%%%%%%%%%%%%%%%
\subsection{Gravitational waves in the Einstein and Jordan frames}
%%%%%%%%%%%%%%%%%%%%%%%%%%%%%%%%%%%%%%%%%%%%%%%%%%%%%%%%%%%%%
In the Jordan frame, we describe perturbations of the metric and of
the scalar field as follows:
\be
g_{\mu\nu}=g_{\mu\nu}^{(0)}+%\epsilon
 h_{\mu\nu}\,,\quad \phi=\phi^{(0)}+%\epsilon
\phi^{(1)}\,.
\ee
In the Einstein frame this corresponds to
$g^{E}_{\mu\nu}=g^{(0)E}_{\mu\nu}+h^E_{\mu\nu}$,
$\varphi=\varphi^{(1)}$,
and from Eqs.~(\ref{JEtransf1})-(\ref{JEtransf3}) one gets
\beq
&&h_{\mu\nu}=F(\phi^{(0)})^{-1}\left(h_{\mu\nu}^{E}-g_{\mu\nu}^{(0)}F'(\phi^{(0)})\phi^{(1)}\right)\,,\label{transform1}\\
&&\phi^{(1)}=\left[\frac{3}{2}\frac{F'(\phi^{(0)})^2}{F(\phi^{(0)})^2}+\frac{8\pi G Z(\phi^{(0)})}{F(\phi^{(0)})}\right]^{-1/2}\varphi^{(1)}\,.\label{transform2}
\eeq
In general, gravitational waves in scalar-tensor theories have 3
degrees of freedom \cite{Eardley:1974nw,Eardley:1973br}. In the
Einstein frame they correspond to the two transverse-traceless
components of the metric perturbation, plus the scalar field.  The
calculation of these quantities does not present difficulties. The
physical degrees of freedom, which should be computed in the Jordan
frame, can be read off from
Eqs.~\eqref{transform1}-\eqref{transform2}.  An alternative procedure
is presented in Ref.~\cite{Barausse:2012da}, which shows the
transformation of the corresponding Newman-Penrose quantities, with
similar results.  In this work we will present results for the scalar
field $\varphi$ and for the curvature scalar $\Psi_4^{E}$ in the
Einstein frame, which is directly related to the two Einstein-frame
polarization states $h_+^{E}$ and $h_\times^{E}$ via
\be
\Psi_4^{E}=\ddot{h}^{E}_+-i\ddot{h}^{E}_{\times}\,,
\ee
where dots denote time derivatives. Quantities in the Jordan frame
can be found using Eqs.~\eqref{transform1}-\eqref{transform2}, once
one specifies the underlying theory. From here onwards, having
established the relation between the Einstein and Jordan frames, we
shall work exclusively in the Einstein frame, dropping the label ``E''
from all quantities. Unless specified otherwise, we will use
geometrical units and set $G=c=1$. Note however that here $G$ is a
bare gravitational constant, and it is different from the quantity
measured by a Cavendish experiment.

%%%%%%%%%%%%%%%%%%%%%%%%%%%%%%%%%%%%%%%%%%%%%%%%%%%%%%%%%%%%%
%%%%%%%%%%%%%%%%%%%%%%%%%%%%%%%%%%%%%%%%%%%%%%%%%%%%%%%%%%%%%
\section{Scalar fields in single and\\binary black-hole backgrounds:\\analytical approximations}
\label{sec:analytic}
%%%%%%%%%%%%%%%%%%%%%%%%%%%%%%%%%%%%%%%%%%%%%%%%%%%%%%%%%%%%%
%%%%%%%%%%%%%%%%%%%%%%%%%%%%%%%%%%%%%%%%%%%%%%%%%%%%%%%%%%%%%

In this Section we introduce 
approximate analytical solutions that describe single and binary BHs
in a scalar field gradient. These solutions will be useful below,
either as code checks or for the interpretation of our numerical
results.

%%%%%%%%%%%%%%%%%%%%%%%%%%%%%%%%%%%%%%%%%%%%%%%%%%%%%%%%%%%%%
\subsection{Single black holes:\\linearized analytical solutions}
\label{sec:singlebh}
%%%%%%%%%%%%%%%%%%%%%%%%%%%%%%%%%%%%%%%%%%%%%%%%%%%%%%%%%%%%%

Let us assume that the scalar-field gradient is of such low amplitude
that the scalar can be treated as a perturbative effect on the
spacetime metric. Under this assumption we can neglect terms quadratic
in the scalar field, and therefore the field equations 
reduce to
\begin{align}
R_{\mu\nu}&=0\,,\label{Eeq1}\\
\Box\varphi&=0\,.\label{scfeq}
\end{align}
We will drop this perturbative approximation in Section
\ref{sec:numerical}, where the field equations
(\ref{Eeq}) and (\ref{Seq}) will be solved numerically.

Equation~(\ref{Eeq1}) is of course identical to Einstein's equations
in vacuum. We will consider the Schwarzschild (and later the Kerr)
metrics as background BH solutions, and we will solve the
Klein-Gordon equation (\ref{scfeq}) on these backgrounds. For the
reasons explained in the introduction we are interested in numerical
evolutions in a scalar-field gradient. Therefore we will consider
background scalar-field solutions generated by distant, fixed,
infinite homogeneous planes with constant surface scalar-charge density
$\sigma$.
To our knowledge, Press \cite{Press:1972} was the first to study a
closely related problem (his setup differs from ours in that he
considered a spherical shell of scalar charge as the source of the
scalar field).
Here we recast some of his results in a form suitable for
comparison with our numerical setup.

%%%%%%%%%%%%%%%%%%%%%%%%%%%%%%%%%%%%%%%%%%%%%%%%%%%%%%%%%%%
\subsubsection{Spherically symmetric black-hole background}
%%%%%%%%%%%%%%%%%%%%%%%%%%%%%%%%%%%%%%%%%%%%%%%%%%%%%%%%%%%

Let us first consider the Schwarzschild solution in isotropic
coordinates:
\begin{eqnarray}
  ds^2 &=& - \frac{\left(1-\frac{M}{2\tr}\right)^2}
                {\left(1+\frac{M}{2\tr}\right)^2}dt^2
           +\left(1+\frac{M}{2\tr}\right)^4 (dx^2+dy^2+dz^2) \nonumber \\
       &=& - \frac{\left(1-\frac{M}{2\tr}\right)^2}
                {\left(1+\frac{M}{2\tr}\right)^2}dt^2
           +\left(1+\frac{M}{2\tr}\right)^4 \nonumber \\
       &&  \times\left[d\tr^2+\tr^2(d\theta^2+\sin^2\theta d\phi^2))
           \right]\,.\label{schwarz}
\end{eqnarray}
Here $\tr=\sqrt{x^2+y^2+z^2}$ is the isotropic radius, which is
related to the areal radius $r$ by
$r=\tr\left(1+\frac{M}{2\tr}\right)^2$.

The scalar-field equation (\ref{scfeq}) on this background admits a
simple axisymmetric solution sourced by distant planes of
constant scalar-charge density, i.e.:
\beq
\varphi_{\rm ext}&=&2\pi\sigma(r-M)\cos\theta=2\pi\sigma\left(\tr+\frac{M^2}{4\tr}\right)\cos\theta\nonumber\\
&=&2\pi\sigma z\left(1+\frac{M^2}{4\tr^{2}}\right)\,,\label{defphiext}
\eeq
where $z=\tr\cos\theta$ is the direction orthogonal to the charged
plane.  If no BH is present, Eq.~(\ref{defphiext}) reduces to
the field $\varphi_{\rm ext}=2\pi \sigma z$ generated by homogeneously
charged infinite planes, with constant gradient $\partial_z
\varphi_{\rm ext}=2\pi\sigma$. For large $|z|$ the constant-gradient
behavior applies also to the case where the background is a
Schwarzschild BH.

We shall take as initial condition $\varphi_{\rm ini}=2\pi \sigma z$
and test our numerical framework by checking that, after a transient,
the scalar-field profile settles to the analytical solution
$\varphi=\varphi_{\rm ext}$, up to corrections of second order in the
scalar field.

In Appendix~\ref{app:boosted} we derive a solution describing a BH
moving with small constant velocity in a direction orthogonal to the
charged planes, and show that it does not emit scalar waves.  Indeed,
as we will show analytically in Section \ref{anbin} and numerically in
Section \ref{sec:numerical}, the BH must have nonvanishing
acceleration in order to generate scalar radiation.

\begin{figure*}[hbt]
\begin{center}
\includegraphics[width=0.47\textwidth]{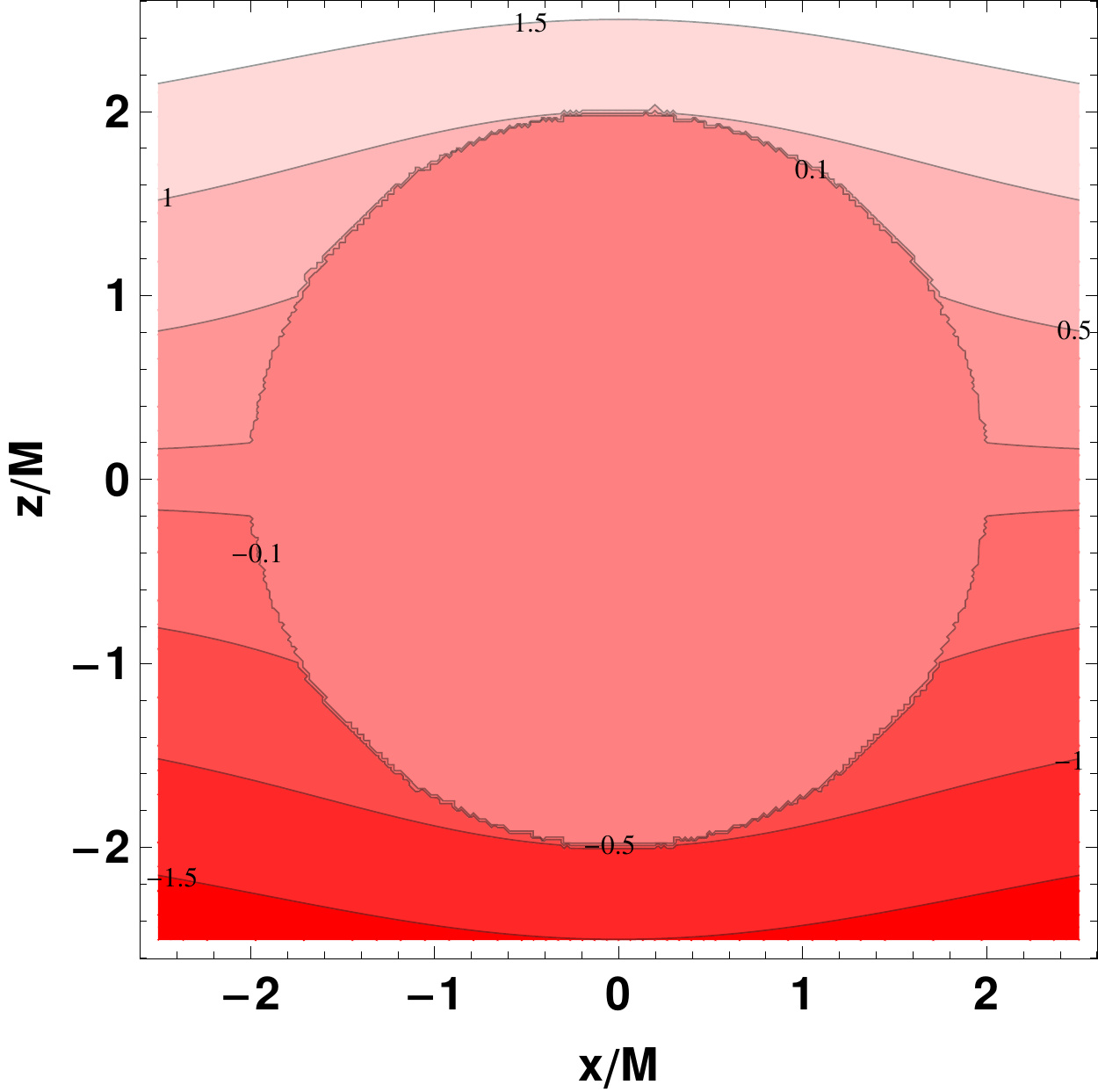}
\includegraphics[width=0.47\textwidth]{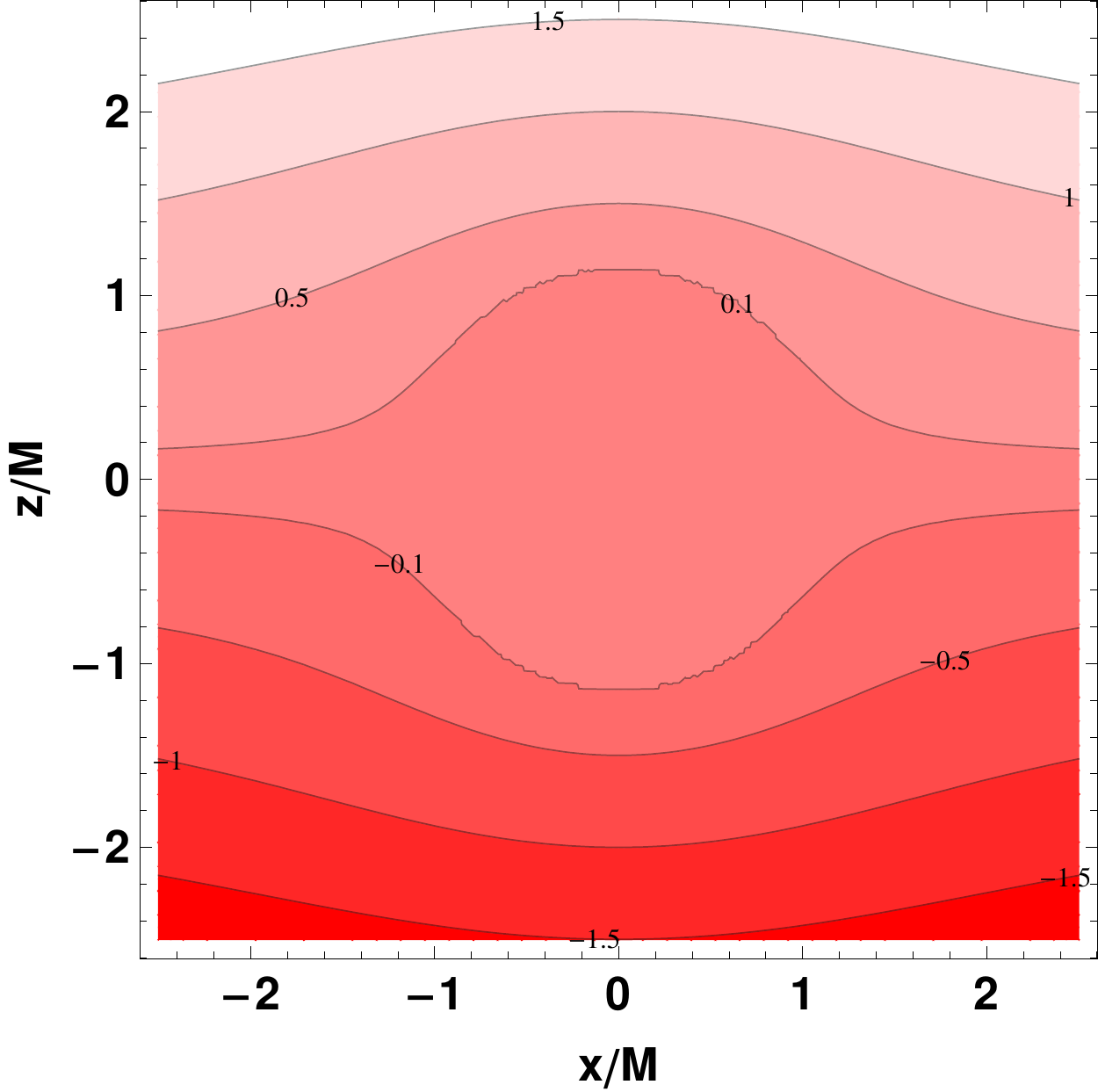}\\
\includegraphics[width=0.47\textwidth]{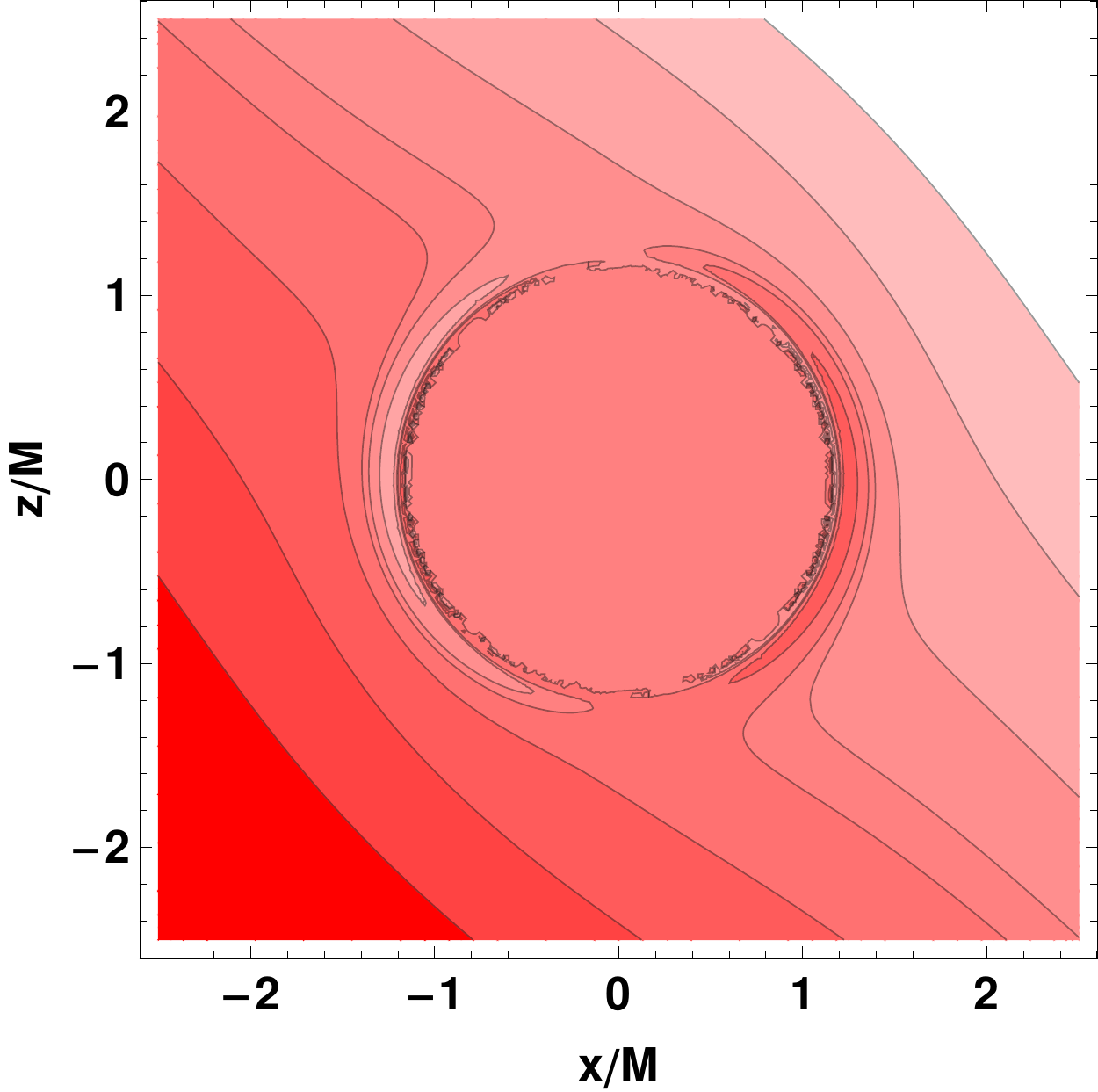}
\includegraphics[width=0.47\textwidth]{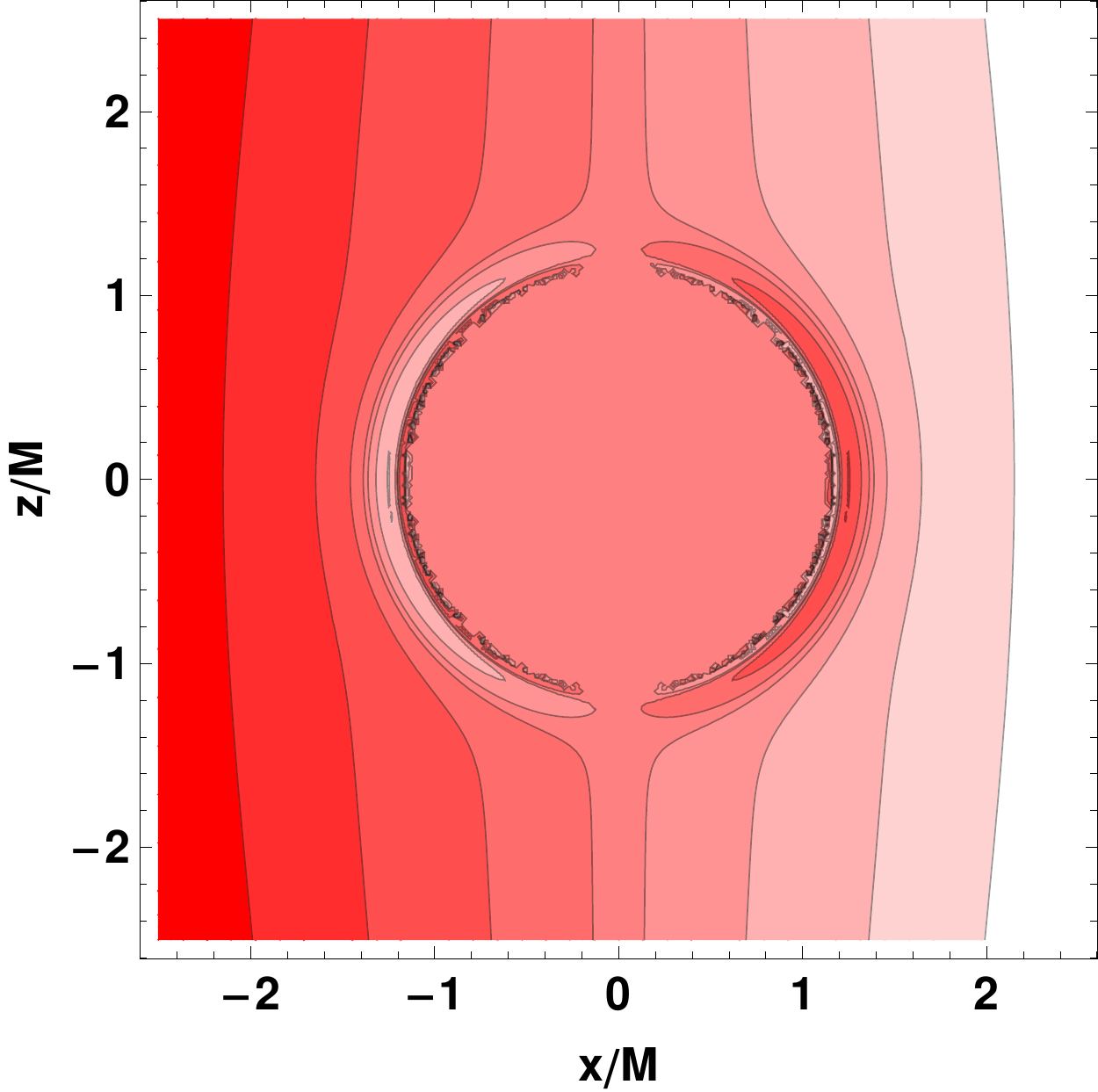}
\end{center}
\caption{Contour plots of the field $\varphi_{\rm ext}$ in the
  vicinity of a rotating BH, as given by
  Eq.~(\ref{rotating_solution}).  Top: The infinite charged plane is
  at an angle $\gamma=0$, and the BH has dimensionless spin $a=0$
  (left) and $a/M=0.99$ (right).  The value of $\varphi_{\rm
    ext}/(2\pi\sigma)$ is shown along selected contour lines; the two
  panels only differ because of the different size of the
  horizon. Bottom: A BH with $a/M=0.99$ is immersed in a field
  gradient at angles $\pi/4$ (left) and $\pi/2$ (right). All contour
  plots refer to the plane $y=0$. Selected contour lines correspond to
  the same values as the top panels.}
\label{fig:contour}
\end{figure*}
%

%%%%%%%%%%%%%%%%%%%%%%%%%%%%%%%%%%%%%%%%%%%%%%%%%%%%%%%%
\subsubsection{Rotating black-hole background}
%%%%%%%%%%%%%%%%%%%%%%%%%%%%%%%%%%%%%%%%%%%%%%%%%%%%%%%%
In the case of a rotating BH with a rotation axis that is not
orthogonal to the charged planes, axisymmetry is lost; however, a
simple solution can still be found \cite{Press:1972}. Let us choose
our coordinates so that the BH spins along the $z$-axis, at an
angle $\gamma$ with respect to the direction orthogonal to the
charge-carrying planes. Then it can be shown that a solution is
\begin{eqnarray}
\varphi_{\rm ext}&=&2\pi\sigma (r-M) \left(\cos\gamma\cos\theta+
\sin\gamma\sin\theta \cos\phi f_a\right) \nn\\
&=&
2\pi\sigma (r-M) \left[\frac{z}{r} \cos\gamma +
\frac{x}{r} f_a \sin\gamma \right]\,,\label{rotating_solution}
\end{eqnarray}
where $\delta=\sqrt{M^2-a^2}$,
\be
f_a=\frac{1}{r-M}{\Re}\left[\delta\frac{\Gamma[2-ia/\delta]}{e^{a\frac{\pi}{2\delta}}}
 P_1^{ia/\delta}\left(\frac{r-M}{\delta}\right)\right] \,,
\ee
$\Gamma$ denotes the $\Gamma$-function and $P_{\mu}^\nu(z)$ is an
associated Legendre function of the first kind.  At large distances
$f_a\to 1$, and it can be seen that the charged plane generates a
uniform gradient in the $xz$-plane.
Figure~\ref{fig:contour} shows contour plots of $\varphi_{\rm ext}$ in
the $y=0$ plane, for different values of $a$ and of the angle
$\gamma$.  Notice how the field lines are distorted and frame-dragged
close to the horizon.

%%%%%%%%%%%%%%%%%%%%%%%%%%%%%%%%%%%%%%%%%%%%%%%%%%%%%%%%%%%%%%%%%%%%%%%%%%%%%%
\subsection{Black-hole binaries: analytical approximation for quasi-circular inspirals}\label{anbin}
%%%%%%%%%%%%%%%%%%%%%%%%%%%%%%%%%%%%%%%%%%%%%%%%%%%%%%%%%%%%%%%%%%%%%%%%%%%%%%

Let us now turn to the more complicated case of a quasi-circular BH binary
evolving in an external scalar field with constant gradient. We begin our
discussion by addressing the delicate problem of specifying boundary conditions
for this system, which will be used to perform the numerical simulations
discussed in Sec.~\ref{sec:numerical}.

%%%%%%%%%%%%%%%%%%%%%%%%%%%%%%%%%%%%%%%%%%%%%%%%%%%%%%%%%%%%%%%%%%%%%%%%%%%%%%
\subsubsection{Boundary conditions}\label{sec:boundary_conditions}
%%%%%%%%%%%%%%%%%%%%%%%%%%%%%%%%%%%%%%%%%%%%%%%%%%%%%%%%%%%%%%%%%%%%%%%%%%%%%%

In our numerical setup the boundary conditions are imposed by
fictitious faraway charges, which are not modeled in our numerical
simulations.  These charges are assumed to lie outside the numerical
grid, and mimic an external profile due (say) to a galactic
scalar-field background.
This ``large-scale field'' acts as a sort of reservoir: when the local
field begins falling into the BH, there is an ingoing flux which
restores the scalar-field gradient. The presence of this field outside
the boundaries of our numerical simulations is implemented through the
boundary conditions. At large distances we want to allow for outgoing
waves while imposing the existence of the scalar-field gradient, so we
require the field to behave in the following way:
\be \label{scalar_boundary_cond}
\varphi=\varphi_{\rm ext}+\frac{\Phi(t-r,\theta,\phi)}{r}\,,
\ee
where $r$ is the areal radius, the constant gradient corresponds to the external field
$\varphi_{\rm ext}=2\pi\sigma(r-M)\cos\theta$, 
and the second term on the right-hand side is the solution of the
homogeneous equation $\Box\varphi=0$ describing outgoing
(approximately spherical) scalar waves.
We thus get
\be
\frac{\partial}{\partial r}\left(r\varphi\right)+
\frac{\partial}{\partial t}\left(r\varphi\right)=\frac{\partial}{\partial r}\left(r\varphi_{\rm ext}\right)\,.
\ee
Since the boundary conditions are defined at large distances, 
$\varphi_{\rm ext}\simeq2\pi\sigma r\cos\theta$, and
we can write
\be
\label{boundary}
\frac{\partial}{\partial r}\left(r\varphi\right)+
\frac{\partial}{\partial t}\left(r\varphi\right)=4\pi\sigma r\cos\theta\,.
\ee
%

%%%%%%%%%%%%%%%%%%%%%%%%%%%%%%%%%%%%%%%%%%%%%%%%%%%%%%%%%%%%%%%%%%%%%%%%%%%%%%
\subsubsection{Multipole expansion of the scalar field}
\label{appsol}
%%%%%%%%%%%%%%%%%%%%%%%%%%%%%%%%%%%%%%%%%%%%%%%%%%%%%%%%%%%%%%%%%%%%%%%%%%%%%%
The angular dependence of the scalar field can be described through a
multipole expansion of the form
\begin{eqnarray}\label{eq:multipole_expansion}
\Phi(t-r,\theta,\phi) = {\cal M} + n^{i} \dot{\cal D}_{i}
+ \frac{1}{2} n^{i}n^{j} \ddot{\cal Q}_{ij} + \cdots \,,\nonumber\\
\end{eqnarray}
where $\Phi$ is the function appearing on the right-hand side of
Eq.~(\ref{scalar_boundary_cond}), dots denote derivatives with respect
to the retarded null coordinate $u=t-r$ and $\vec{n} = \vec{x}/r =
\vec{x}/|\vec{x}|$ is the radial unit vector, which depends only on
the angles $(\theta,\phi)$. The calligraphic symbols ${\cal M}$,
${\cal D}_i$ and ${\cal Q}_{ij}$ denote the monopolar, dipolar and
quadrupolar components of $\Phi$, respectively.

The full relativistic scalar equation to be solved is $\Box \varphi =
0$. In order to obtain a solution which describes the physics that we
are interested in -- namely, a BH binary in a scalar gradient -- it is
essential to impose correct boundary conditions, both at null
infinity, and in the vicinity of the worldlines of the singularities
of the two BHs. The boundary conditions at null infinity have been
discussed above; the near-worldline boundary conditions that must be
imposed in order to find an approximate solution are more complicated.

In the special case of a comparable-mass BH binary system with small
size-to-separation ratio (or, equivalently, small orbital velocity)
the problem of imposing correct near-worldline boundary conditions is
substantially simplified when one employs a ``point-particle''
effective field theory, in which length scales smaller than the
Schwarzschild radii are integrated out.  In this effective field
theory the matter action has the form
\begin{equation}
\label{eq:pp_action}
S_{\rm matt}^{\rm pp} = \sum_{A} \int_{\Gamma_{A}} ds_{A} \mathcal{L}_{A} \,,
\end{equation}
where $A$ is an index that runs over the bodies ($A=1,2$ for a binary
system), $\Gamma_{A}$ is the worldline of body $A$, $ds_{A}$ is the
proper differential arclength along $\Gamma_{A}$, and
$\mathcal{L}_{A}$ is the ``effective point-particle Lagrangian'' of
body A. For a structureless particle of mass $m_{A}$, one has
$\mathcal{L}_{A} = -m_{A}$.

The matter action (\ref{eq:pp_action}) gives rise to sources in the
field equations of the effective point-particle theory, and these
sources automatically enforce the correct boundary conditions at the
worldlines $\Gamma_{A}$ for both $g_{\mu \nu}$ and $\varphi$. For
instance, to leading nonrelativistic order, the scalar-field equation
has the explicit form
\begin{eqnarray}
\Box_f \varphi &=& 
4 \pi \rho_{\varphi} \label{weqflat}
\nonumber
\\
&=& 
4 \pi \sum_{A=1}^{2} Q_{A} \, \delta^{(3)}(\vec{x}-\vec{z}_{A}(t)) \,,
\end{eqnarray}
where $\Box_f$ is the D'Alembert operator in flat space,
$(t,\vec{z}_{A}(t))$ is an explicit parametrization of the worldline
$\Gamma_{A}$, and $Q_{A}$ are the scalar charges of the BHs.

Moreover, to leading nonrelativistic order, one finds that the
multipole moments entering into Eq.~(\ref{eq:multipole_expansion}) are
given by
\begin{eqnarray}\label{scalar_mon_mom}
{\cal M} &=&
 \int d^{3}x \, \rho_{\varphi} 
= 
 Q_{1} + Q_{2}\,,
\\
\label{scalar_dip_mom}
\vec{\cal D} &=& 
 \int d^{3}x \, \vec{x} \, \rho_{\varphi} 
=
 Q_{1} \vec{z}_{1} + Q_{2} \vec{z}_{2} \,,
\end{eqnarray}
and so on.

In general, calculating the scalar charges $Q_{A}$ is a difficult
problem. A simplification takes place if we assume that the BH masses
$M_{A}$ have the same order of magnitude $M$, and that
\begin{equation}\label{small_scalar_gradient}
\sigma \ll 
\sqrt{\frac{a}{M^{3}}} \sim \frac{1}{Mv} \,,
\end{equation}
where $a$ is the typical orbital separation, and $v$ is the typical
orbital velocity.  Henceforth, terms of order 
$(Mv\sigma)^{2}$ will be dropped.  Then the BH scalar charges $Q_{A}$
may be found by Jacobson's formula \cite{Jacobson:1999vr}, which for
Schwarzschild BHs yields
\begin{eqnarray}
\label{ScalarCharge}
Q_{A}(t) &=& 
4M_{A}^{2} \, \left[\frac{\partial \varphi(t,\vec{z}_{A}(t)) }{\partial t} + \vec{v}_{A}(t) \cdot \vec{\nabla} \varphi(t,\vec{z}_{A}(t)) \right]
\nonumber
\\
&=& 
8 \pi \sigma M_{A}^{2} \, \vec{v}_{A}(t) \cdot \hat{z} \,,\label{Qgrad}
\end{eqnarray}
where $\vec{v}_{A}(t) = \dot{\vec{z}}_{A}(t)$ is the velocity of body
$A$, and in the second line the full scalar field $\varphi(t,\vec{x})$
has been replaced by the zeroth-order field $\varphi_{\rm ext} = 2 \pi
\sigma z$.

Let us specialize to quasi-circular orbits in the $yz$-plane, so that
the trajectories take the simple form
\begin{eqnarray}
\vec{z}_{\rm rel}(t) &=& \vec{z}_{1}(t) - \vec{z}_{2}(t) 
\nonumber
\\
&=& a(t) [\hat{y} \cos \chi (t) + \hat{z} \sin \chi (t)] \,,
\\
M\vec{z}_{\rm CM}(t) &=& M_{1} \vec{z}_{1}(t) + M_{2} \vec{z}_{2}(t)  \,,
\end{eqnarray}
where $a(t)$ is the orbital radius, $\chi (t) = \int \omega (t) dt$ is
the orbital phase, $\omega(t)$ is the angular frequency of the orbit,
and $\vec{z}_{\rm CM}(t)$ is the center of mass of the binary system.
The quantities $\dot{a}$, $\ddot{\chi}=\dot{\omega}$, and
$\vec{a}_{\rm CM} = \ddot{\vec{z}}_{\rm CM}$ are all small (of order
$1/c^{5}$), and vanish in the absence of radiation reaction. An
explicit expression for their leading-order time evolution may be
found by solving the 2.5PN equations of motion (given in Section 9 of
\cite{Blanchet:2002av}), while dropping conservative
corrections. Carrying out this calculation yields
\begin{eqnarray}
a(t) &=& 
a^{(0)} \left( 1 - \frac{256t}{5\tau_{\rm q}}\right)^{1/4} \simeq
a^{(0)} \left( 1 - \frac{64t}{5\tau_{\rm q}}\right) \,,
\\
\dot{\chi}(t) &=& 
\dot{\chi}^{(0)} \left( 1 - \frac{256t}{5\tau_{\rm q}} \right)^{-3/8} \simeq
\dot{\chi}^{(0)} \left( 1 + \frac{96t}{5\tau_{\rm q}} \right) \,,\label{evolution_eqs}
\end{eqnarray}
where $a^{(0)}$ and $\dot{\chi}^{(0)}$ are the (constant) radius and
angular frequency of the zeroth-order orbit, respectively, and
\begin{equation}
\tau_{\rm q} = 
\left[ \frac{M_{1}M_{2}(M_{1}+M_{2})}{[a^{(0)}]^{4}} \right]^{-1}
\end{equation}
is the time scale over which the quadrupole tensor radiation shrinks
the orbit. 

With an explicit description of the orbit in hand, the scalar charges
$Q_{1,2}$ and multipole moments
(\ref{scalar_mon_mom})-(\ref{scalar_dip_mom}) may be calculated (see
Appendix \ref{app:binaryscaleq} for the explicit expressions). In this
way, we find that: 

\begin{itemize}
\item[1)] monopole radiation is emitted at the orbital frequency:
\beq
{\cal M} &=&
\frac{8 \pi \sigma}{M} 
\biggl[ M(M_{1}^{2} + M_{2}^{2})(\vec{v}_{\rm CM} \cdot \hat{z})\\
&+& M_{1}M_{2}(M_{1}-M_{2})(\dot{a} \sin \chi + a \dot{\chi} \cos \chi) \biggr] \,,
\nn
\eeq
and it vanishes in the equal-mass limit;
\item[2)] dipole radiation is emitted at {\it twice} the orbital
  frequency, and more precisely
\be
\dot{\vec{\cal D}} = 
\dot{\vec{\cal D}}_{\rm CM} + 
\dot{\vec{\cal D}}_{\rm rel,\, DC} + 
\dot{\vec{\cal D}}_{\rm rel,\, osc} \,,
\ee
where the ``CM'' term is emitted at the orbital frequency,
the ``DC'' component is nonoscillatory, and the $\vec{\cal D}_{\rm
  rel,\, osc}$ component oscillates at twice the orbital frequency:
cf.~Eq.(\ref{app:dipoledot_eq}).
\end{itemize}

The physical problem addressed here differs from the situation
investigated in \cite{Horbatsch:2011ye}. That study considered
time-dependent scalar boundary conditions, rather than a gradient, and
it found that monopole radiation is absent, while dipole radiation
vanishes in the equal-mass limit. One may expect dipole scalar
radiation to be emitted at the orbital frequency, rather than {\em
  twice} the orbital frequency. The reason why this expectation is
erroneous in our case is that we have a background field with a
gradient directed along the orbital plane, which combines with the
oscillatory component sourced by the orbital motion.

A simple toy model can provide us with a complementary and perhaps
more intuitive way to justify the expectation that dipole radiation
must be emitted at twice the orbital frequency. Let us consider a
rotating source with frequency $\Omega$ on a scalar-field background
$\varphi_{\rm ext}=2\pi\sigma z=2\pi\sigma r\sin\theta\sin\phi$ [in
  our ``rotated'' polar coordinates, see Eqs.~(\ref{noncanonical})
  below]. The source will produce a modulation in the background field
of the form
\begin{equation}
\varphi=\varphi_{\rm ext}[1+f(\phi-\Omega t)]\,.\label{oneplus}
\end{equation} 
Expanding in circular harmonics, $f(\phi-\Omega t)=\sum_m f_me^{\ii
  m(\phi-\Omega t)}$ and
\begin{equation}
\varphi=2\pi\sigma r\sin\theta\sin\phi(1+\sum_mf_me^{\ii m(\phi-\Omega t)})\,,
\end{equation}
which implies that the multipolar components of the field will have
the following dependence:
\begin{equation}
\varphi_{lm}\sim (e^{-\ii(m+1)\Omega t}+e^{-\ii(m-1)\Omega t})+{\rm constant}\,.
\end{equation}
Therefore the $m=0$ contribution should oscillate with frequency
$\Omega$, the $m=1$ contribution with frequency $2\Omega$, the $m=2$
contribution with frequencies $3\Omega$ and $\Omega$, and so on. As we
will show below, this behavior is consistent with our numerical
simulations.

%%%%%%%%%%%%%%%%%%%%%%%%%%%%%%%%%%%%%%%%%%%%%%%%%%%%%%%%%%%%%
%%%%%%%%%%%%%%%%%%%%%%%%%%%%%%%%%%%%%%%%%%%%%%%%%%%%%%%%%%%%%
\section{Numerical implementation}
\label{sec:numerical}
%%%%%%%%%%%%%%%%%%%%%%%%%%%%%%%%%%%%%%%%%%%%%%%%%%%%%%%%%%%%%
%%%%%%%%%%%%%%%%%%%%%%%%%%%%%%%%%%%%%%%%%%%%%%%%%%%%%%%%%%%%%

Our numerical implementation of scalar-tensor theory closely parallels
\cite{Salgado:2008xh}, but borrowing notation and conventions from
\cite{Zilhao:2010sr}. The physical system studied in
\cite{Zilhao:2010sr} was quite different, since that paper considered
higher-dimensional Einstein gravity in vacuum. However the equations
can be cast as a system involving a scalar field coupled to gravity
via dimensional reduction, so they are formally similar to the system
considered here, as we show below.

%%%%%%%%%%%%%%%%%%%%%%%%%%%%%%%%%%%%%%%%%%%%%%%%%%%%%%%%%%%%%
\subsection{$3+1$ decomposition}
%%%%%%%%%%%%%%%%%%%%%%%%%%%%%%%%%%%%%%%%%%%%%%%%%%%%%%%%%%%%%

As a preliminary step for our numerical implementation, we perform a
$3+1$ decomposition of the spacetime (see \cite{Alcubierre:2008} and
references therein). Let us consider a slicing of the spacetime in a
set of three-dimensional surfaces $\Sigma$. Introducing the normal
$n_\mu$ to the surface $\Sigma$ and the projector
\begin{equation}
\gamma_{\mu\nu}=g_{\mu\nu}+n_\mu n_\nu\,,
\end{equation}
we write the four-dimensional metric in the form ($\mu,\nu=0,\dots,3$;
$i,j=1,2,3$):
\begin{align}
ds^2&=g_{\mu\nu}dx^\mu dx^\nu\nonumber\\
&=-\alpha^2 dt^2+\gamma_{ij}(dx^i+\beta^idt)(dx^j+\beta^jdt)\,,
\end{align}
where $\alpha,\beta^i$ are the lapse and the shift, respectively, and
\begin{equation}
\partial_t=\alpha n+\beta\,.
\end{equation}
We shall denote by $D_i$ the covariant derivative on $\Sigma$, i.e.,
the covariant derivative with respect to the three-dimensional metric
$\gamma_{ij}$.

The Lie derivative with respect to $n^\mu$, then, is ${\cal
  L}_n=(\partial_t-{\cal L}_\beta)/\alpha$. Defining the extrinsic
curvature
\begin{equation}
K_{ij}\equiv-\frac{1}{2}{\cal L}_n\gamma_{ij}\,,
\end{equation}
we have the following evolution equations for the (three-dimensional)
metric:
\begin{equation}
(\partial_t-{\cal L}_\beta)\gamma_{ij}=-2\alpha K_{ij}\,.\label{evolgamma}
\end{equation} 
Furthermore, we define the scalar curvature $K_\varphi$ as
\begin{equation}
K_\varphi=-\frac{1}{2}{\cal L}_n\varphi
\label{eq:Kphi}
\end{equation}
so that
\begin{equation}
(\partial_t-{\cal L}_\beta)\varphi=-2\alpha K_\varphi\,.\label{evolphi}
\end{equation}
Comparing with the definitions of the variables $Q_i$ and $\Pi$ in
\cite{Salgado:2008xh}, we find that
\begin{align}
Q_i&=\frac{D_i\varphi}{\sqrt{8\pi G}}=\frac{\gamma_{i\mu}\partial^\mu\varphi}{\sqrt{8\pi G}}\,,\nonumber\\
\Pi&=\frac{{\cal L}_n\varphi}{\sqrt{8\pi G}}\,.
\end{align}
Therefore,
\begin{equation}
K_\varphi=-\frac{1}{2} \sqrt{8\pi G}\Pi\,.
\end{equation}
The constraint equations (2.15) and (2.16) of \cite{Salgado:2008xh} read
\begin{align}
^{(3)}R+K^2-K_{ij}K^{ij}&=\frac{\Pi^2+Q^2}{f}=4K_\varphi^2+
D_i\varphi D^i\varphi\,,
\label{constraintR}\\
D_lK^l_{~i}-D_iK&=-\frac{\Pi Q_i}{f}=2K_\varphi D_i\varphi\,.\label{constraintK}
\end{align}

The evolution equation (2.17) of \cite{Salgado:2008xh} can be written
as
\begin{align}
&(\partial_t-{\cal L}_\beta)K_{ij}
=(\partial_t-{\cal
  L}_\beta)\gamma_{ik}K^k_{~j}\nonumber\\
&=-D_iD_j\alpha+\alpha\left(^{(3)}R_{ij}+KK_{ij}-
D_i\varphi D_j\varphi-2K_{ik}K^k_{~j}\right)\,.\label{evolKij}
\end{align}
Note that since  $\gamma^\mu_{~i} \gamma^\nu_{~j}R_{\mu\nu}=D_i\varphi
D_j\varphi$, this expression coincides with Eq.~(2.23) of \cite{Zilhao:2010sr}.
Taking the trace of Eq.~(\ref{evolKij}) we have
\begin{equation}
\partial_tK-\beta^l\partial_lK+D^iD_i\alpha-\alpha\left({}^{(3)}R+K^2-D^i\varphi
D_i\varphi\right)\,,
\end{equation}
and using Eq.~(\ref{constraintR}) we find 
\begin{equation}
(\partial_t-{\cal L}_\beta)K=-D^iD_i\alpha+\alpha K_{ij}K^{ij}+4\alpha K_\varphi^2\,,
\label{evolK}
\end{equation}
that should be compared to Eq.~(2.19) of \cite{Salgado:2008xh}.  The
evolution equation (2.18) of \cite{Salgado:2008xh} can be written as
\begin{align}
&\frac{1}{\alpha}(\partial_t-{\cal L}_\beta)K_\varphi=-\frac{1}{2\sqrt{8\pi G}}
{\cal L}_n\Pi\nonumber\\
&=K_\varphi K-\frac{1}{2\alpha}D^i\varphi D_i\alpha-\frac{1}{2}D_iD^i\varphi\,.
\label{evolKphi}
\end{align}
All terms of this expression appear, with the same coefficients, in
Eq.~(2.37) of \cite{Zilhao:2010sr}. The additional terms in that
equation which are not present here are due to the more complicated
dynamics of the scalar field arising from dimensional reduction.

It can be useful to write the equations also in terms of the
stress-energy tensor, which enables us to compare the scalar-field
terms with those in \cite{Alcubierre:2008}. From (\ref{Eeq}) we have
\begin{equation}
8\pi GT_{\mu\nu}=\partial_\mu\varphi\partial_\nu\varphi-\frac{1}{2}g_{\mu\nu}
\partial_\alpha\varphi\partial^\alpha\varphi\,.
\end{equation}
Since $n^\mu\partial_\mu\varphi=-2K_\varphi$,
$\gamma^{\mu\nu}\partial_\mu\varphi=D^\nu\varphi$,
$g_{\mu\nu}=\gamma_{\mu\nu}-n_\mu n_\nu$, defining, as on page 87 of
\cite{Alcubierre:2008}
\begin{align}
\rho&=n^\mu n^\nu T_{\mu\nu}\,,\\
j^\alpha&=-\gamma^{\alpha\mu}n^\nu T_{\mu\nu}\,,
\end{align}
we get
\begin{align}
8\pi G~\rho&=n^\mu n^\nu\partial_\mu\varphi\partial_\nu\varphi-\frac{1}{2}
g_{\mu\nu}\partial_\alpha\varphi\partial^\alpha\varphi\\
&=2K_\varphi^2+\frac{1}{2}D_i\varphi D^i\varphi\,,\\
8\pi G~j^i&=-8\pi G\gamma^{i\mu}n^\nu T_{\mu\nu}\\
&=-D^i\varphi n^\mu\partial_\mu\varphi=2K_\varphi D^i\varphi\,,
\end{align}
where we have used the fact that $\gamma^{00}=0$ (see
\cite{Alcubierre:2008}),
thus $\gamma^{\mu\nu}D_\mu\varphi
D_\nu\varphi=\gamma^{ij}D_i\varphi D_j\varphi$.  Therefore,
Eqs.~(2.4.6) and (2.4.9) coincide with our constraint equations
(\ref{constraintR}) and (\ref{constraintK}).
Furthermore, we can compute the quantity (cf. page 89 of
\cite{Alcubierre:2008})
\begin{equation}
S^{\mu\nu}=\gamma^{\mu\alpha}\gamma^{\nu\beta}T_{\alpha\beta}\,.
\end{equation}
We have
\begin{equation}
8\pi GS^{\mu\nu}
=D^\mu\varphi D^\nu\varphi-\frac{1}{2}\gamma^{\mu\nu}D^i\varphi D_i\varphi
+2\gamma^{\mu\nu}K_\varphi^2\,,
\end{equation}
and the trace of this equation (since $\gamma^\mu_{~\mu}=3$) yields
\begin{equation}
8\pi GS=-\frac{1}{2}D^i\varphi D_i\varphi+6K_\varphi^2\,.
\end{equation}
Then, $8\pi G(S-\rho)=4K_\varphi^2-D_i\varphi D^i\varphi$, and 
\begin{equation}
4\pi G\left[(S-\rho)\gamma_{ij}-2S_{ij}\right]=-D_i\varphi D_j\varphi\,,
\end{equation}
therefore Eq.~(2.5.6) of \cite{Alcubierre:2008} coincides with our
Eq.~(\ref{evolKij}).

%%%%%%%%%%%%%%%%%%%%%%%%%%%%%%%%%%%%%%%%%%%%%%%%%%%%%%%%%%%%%
\subsection{Baumgarte-Shapiro-Shibata-Nakamura formalism}
%%%%%%%%%%%%%%%%%%%%%%%%%%%%%%%%%%%%%%%%%%%%%%%%%%%%%%%%%%%%%
Our evolution equations use the Baumgarte-Shapiro-Shibata-Nakamura
(BSSN) formalism, in which the dynamical variables are
$\{\chi,\gamma_{ij},\tilde A_{ij},K,{\tilde\Gamma}^i\}$, defined as
follows:
\begin{align}
&\gamma_{ij}=\chi^{-1}\tilde\gamma_{ij}\quad ({\rm with}~\gamma^{ij}=\chi\tilde\gamma^{ij})\,,\nonumber\\
&\chi=(\det\gamma_{ij})^{-1/3}\,,\nonumber\\
&{\tilde A}_{ij}=\chi\left(K_{ij}-\frac{1}{3}\gamma_{ij}K\right)\,,\nonumber\\
&\Gamma^k=\gamma^{ij}\Gamma^k_{ij}=\chi{\tilde\Gamma}^k+\frac{1}{2}{\tilde\gamma}^{kj}
\partial_i\chi\,.
\label{eq:BSSNvars}
\end{align}
Alternative notations replace our variable $\chi$ by a variable $\psi$
defined as $\chi^{-1}=\psi^4$. Here we will use $\chi$ as our
dynamical quantity.

The Einstein equations in the BSSN formulation in the presence of a
scalar field, which appears through the quantities $\rho,j^i,S_{ij}$,
are implemented in an extended version of the {\sc Cactus}
\cite{Cactusweb} based {\sc Lean} code
\cite{Sperhake:2006cy,Sperhake2010a} in the form given in
Eqs.~(\ref{eq:gammat})-(\ref{eq:eta_beta}) of Appendix
\ref{sec:evolution_equations}. Mesh refinement for our simulations is
provided by {\sc Carpet} \cite{Schnetter:2003rb}, horizon diagnostics
by {\sc AHFinderDirect} \cite{Thornburg1996,Thornburg2004} and BH
binary initial data satisfying Eq.~(\ref{Eeq1}) by a spectral solver
\cite{Ansorg:2004ds} provided through {\sc Cactus} as the {\sc
  TwoPunctures} thorn.

%%%%%%%%%%%%%%%%%%%%%%%%%%%%%%%%%%%%%%%%%%%%%%%%%%%%%%%%%%%%%%%%%%%%%%%%%%%%%%
\subsection{Initial data}
%%%%%%%%%%%%%%%%%%%%%%%%%%%%%%%%%%%%%%%%%%%%%%%%%%%%%%%%%%%%%%%%%%%%%%%%%%%%%%

Our initial data sets consist of either a single BH or a BH binary
evolving in a background scalar field with a nonvanishing
gradient. The background scalar field is generated by distant sources,
which are kept fixed in our time evolutions. As a simple way to
enforce this scenario, imagine that the background scalar field is
generated by infinite homogeneous charged planes with surface density
$\sigma$. As we saw in Section \ref{sec:analytic}, if a BH is present
in the spacetime and the scalar field is small enough to be treated in
the linear approximation the metric is unaffected, but the equilibrium
solution for the scalar field changes. This setup is quite similar in
spirit to the one adopted by Palenzuela and collaborators
\cite{Palenzuela:2009hx,Palenzuela:2009yr}, the main difference being
that they dealt with electromagnetic fields, and that their external
magnetic field is generated by a current loop far away from the
system.

Under these assumptions, the solution $\varphi_{\rm ext}$ we found in
Eq.~(\ref{defphiext})
should be a good approximation to the initial data describing a
single, nonrotating BH in a scalar-field gradient. Except for regions
close to the BH singularity, the term $M^2/4\hat{r}^2$ in parentheses
in Eq.~(\ref{defphiext}) can furthermore be neglected. Therefore we
initialize the scalar field using the simplified expression
$\varphi_{\rm ext}=2\pi \sigma z$.
Our numerical simulations of single BHs show indeed that the initial
data (\ref{defphiext}) and its approximate version $2\pi\sigma z$
yield (after a brief transient, which we exclude from our analysis
below) virtually identical evolutions of the scalar field and of the
spacetime metric.

%%%%%%%%%%%%%%%%%%%%%%%%%%%%%%%%%%%%%%%%%%%%%%%%%%%%%%%%%%%%%%%%%%%%%%%%%%%%%%
\subsection{Upper bounds on the field gradient from the threshold of black-hole formation}
\label{upperbound}
%%%%%%%%%%%%%%%%%%%%%%%%%%%%%%%%%%%%%%%%%%%%%%%%%%%%%%%%%%%%%%%%%%%%%%%%%%%%%%
%
Our setup consists of a constant scalar-field gradient at large
distances. Because the energy density in any classical field theory is
proportional to the square of the field gradient, one expects a
roughly constant energy density $\rho \sim \sigma^2$. The total mass
$M$ in a region of linear dimension $R$ then scales like $\sim \rho
R^3$, and therefore $M/R \sim \rho R^2\sim \sigma^2 R^2$. Therefore we
expect that the initial data will contain a horizon for $\sigma
\gtrsim R^{-1}$. This condition imposes a nontrivial constraint on the
size of our numerical grid. Here we provide a more formal argument
supporting this conclusion.

We focus on conformally flat backgrounds with $\tA^j{}_i=0$,
$\Pi=0$. Then the momentum constraint is identically satisfied, while
the Hamiltonian constraint yields
\begin{equation}
  0= \bigtriangleup\psi +\f{1}{8}\psi\eta^{ij}\partial_i\varphi\partial_j\varphi\,,
\end{equation}
where $\eta^{ij}$ is the Minkowski metric.  A scalar field with
constant gradient is such that $\varphi=2\pi\sigma z$, and
therefore we get the equation
\be
\bigtriangleup\psi =-\frac{\pi^2\sigma^2}{2}\psi\,.
\ee
Imposing regularity at $r=0$, the solution of this equation is
\be
\psi=\frac{A\sin{\left(\pi\sigma r/\sqrt{2}\right)}}{r}\,.
\ee
An apparent horizon exists for $(\psi^4r^2)'=0$. This condition is
equivalent to finding the roots of
\be
x\cot{x}=1/2\,,\quad {\rm with} \quad x\equiv \frac{\pi\sigma r}{\sqrt{2}}\,.
\ee
The smallest root of this equation is at $x \simeq 1.16556$, i.e.
\be
\sigma  \simeq 1.16556\frac{\sqrt{2}}{\pi\,r} \simeq \frac{0.52469}{r}\,,
\ee
in good agreement with the previous back-of-the-envelope estimate.

Our boundary conditions are enforced at distance $r/M = 160$ and $384$,
respectively, for single and binary BH simulations.
This provides a formal upper bound of $M\sigma\approx 5\times 10^{-3}$
on the magnitude of the gradients that we can simulate. However, as we
will see below, even smaller gradients $M\sigma \approx 10^{-4}$ can
generate exponentially growing instabilities (indicating collapse) in
our numerical evolutions.

%%%%%%%%%%%%%%%%%%%%%%%%%%%%%%%%%%%%%%%%%%%%%%%%%%%%%%%%%%%%%%%%%%%%%%%%%%%%%%
%%%%%%%%%%%%%%%%%%%%%%%%%%%%%%%%%%%%%%%%%%%%%%%%%%%%%%%%%%%%%%%%%%%%%%%%%%%%%%
\section{Numerical results}
\label{sec:results}
%%%%%%%%%%%%%%%%%%%%%%%%%%%%%%%%%%%%%%%%%%%%%%%%%%%%%%%%%%%%%%%%%%%%%%%%%%%%%%
%%%%%%%%%%%%%%%%%%%%%%%%%%%%%%%%%%%%%%%%%%%%%%%%%%%%%%%%%%%%%%%%%%%%%%%%%%%%%%

In this Section we first discuss our results for single BH evolutions,
verifying that for small values of the gradient $M\sigma$ they are in
good agreement with the analytical predictions of Section
\ref{sec:singlebh}. Then we study gravitational and scalar radiation
from BH binaries in a scalar-field gradient.

%%%%%%%%%%%%%%%%%%%%%%%%%%%%%%%%%%%%%%%%%%%%%%%%%%%%%%%%%%%%%%%%%%%%%%%%%
\subsection{Single black-hole evolutions}
%%%%%%%%%%%%%%%%%%%%%%%%%%%%%%%%%%%%%%%%%%%%%%%%%%%%%%%%%%%%%%%%%%%%%%%%%

In order to compare our numerical results with the analytical
predictions of Section \ref{sec:analytic}, it is important to minimize
coordinate effects.  Let us denote by $\tilde r$ the radial coordinate
used in our numerical simulation, which coincides
with the isotropic coordinate at $t=0$, i.e.
$\tilde r(t=0)=\tr$.
This need not be true at later times, since the gauge can dynamically
change during the evolution.
In order to monitor the scalar field as a function of time and check
that it eventually settles to a configuration close to the analytical
solution $\varphi_{\rm ext}$ of Eq.~(\ref{defphiext}), we need to
perform integrations over spheres at given values of $\tilde r$.
Furthermore, we compute the areal radii $r$ at these locations by
performing spherical integrations of the metric components, as
discussed in Ref.~\cite{Witek:2010xi}.

\begin{figure}[t]
\begin{center}
\includegraphics[width=0.48\textwidth]{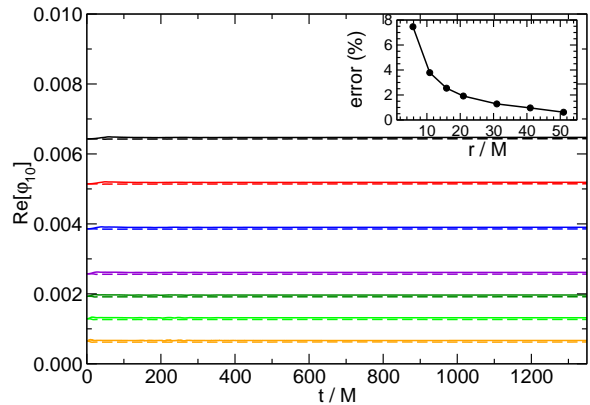}
\end{center}
\caption{
  Real part of the scalar dipole mode ${\varphi}_{10}$ (the imaginary part
  vanishes) for $M\sigma=10^{-5}$ and extraction radii (from top to bottom)
  $\tilde{r}/M=50$, $40$, $30$, $20$, $15$, $10$ and $5$,
  compared to the predictions of
  Eq.~(\ref{an_prediction}). Solid lines refer to the numerical evolution;
  dashed lines refer to the analytical solution evaluated at time-dependent
  areal radii $r$, which are computed dynamically during the evolution. The
  inset shows the percentage discrepancy between the numerical and analytical
  prediction as a function of extraction radius.}
\label{fig:phi_sigma_allradii}
\end{figure}
\begin{figure}[thb]
\begin{center}
\includegraphics[width=0.48\textwidth]{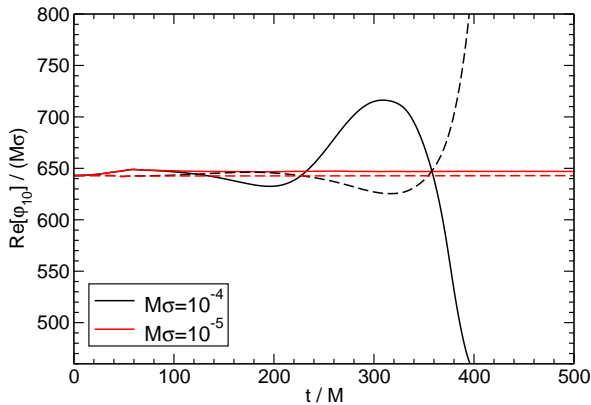}
\end{center}
\caption{Real part of the scalar dipole mode ${\varphi}_{10}$
  (rescaled by $M\sigma$) at the largest extraction radius
  $\tilde{r}/M=50$ for $M\sigma=10^{-4}$ and $M\sigma=10^{-5}$,
  compared to the predictions of Eq.~(\ref{an_prediction}).  Solid
  lines refer to the numerical evolution; dashed lines refer to the
  analytical solution evaluated at time-dependent areal radii $r$,
  which are computed dynamically during the evolution.  The evolution
  does not settle to the analytical solution for $M\sigma=10^{-4}$:
  there is an exponentially growing mode. This also shows up as an
  exponential growth of the subleading multipoles, as can be seen in
  Fig.~\ref{fig:phi_sigma}.}
\label{fig:phi_sigma45}
\end{figure}
\begin{figure*}[hbt]
\begin{center}
\includegraphics[width=0.48\textwidth]{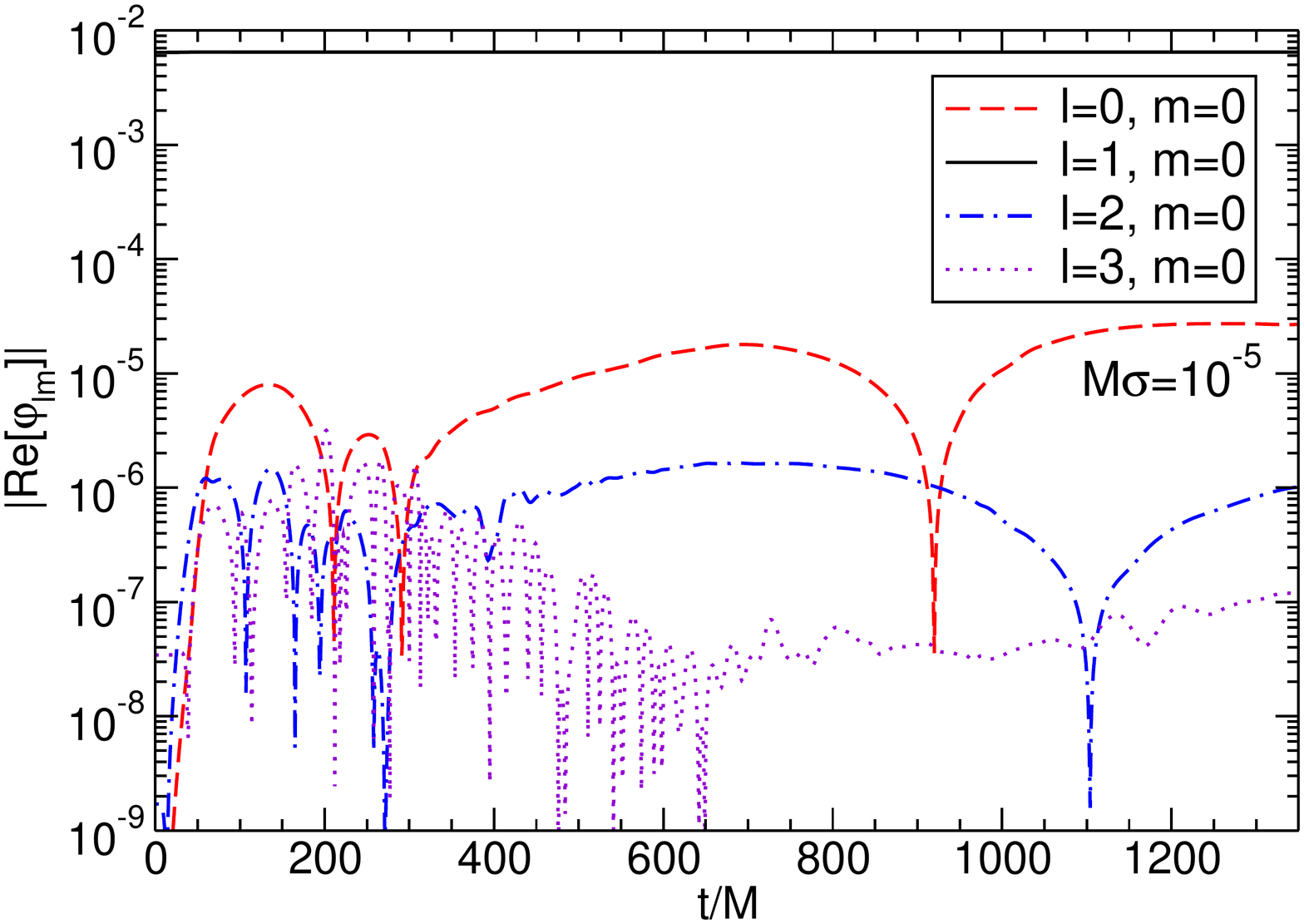}
\includegraphics[width=0.48\textwidth]{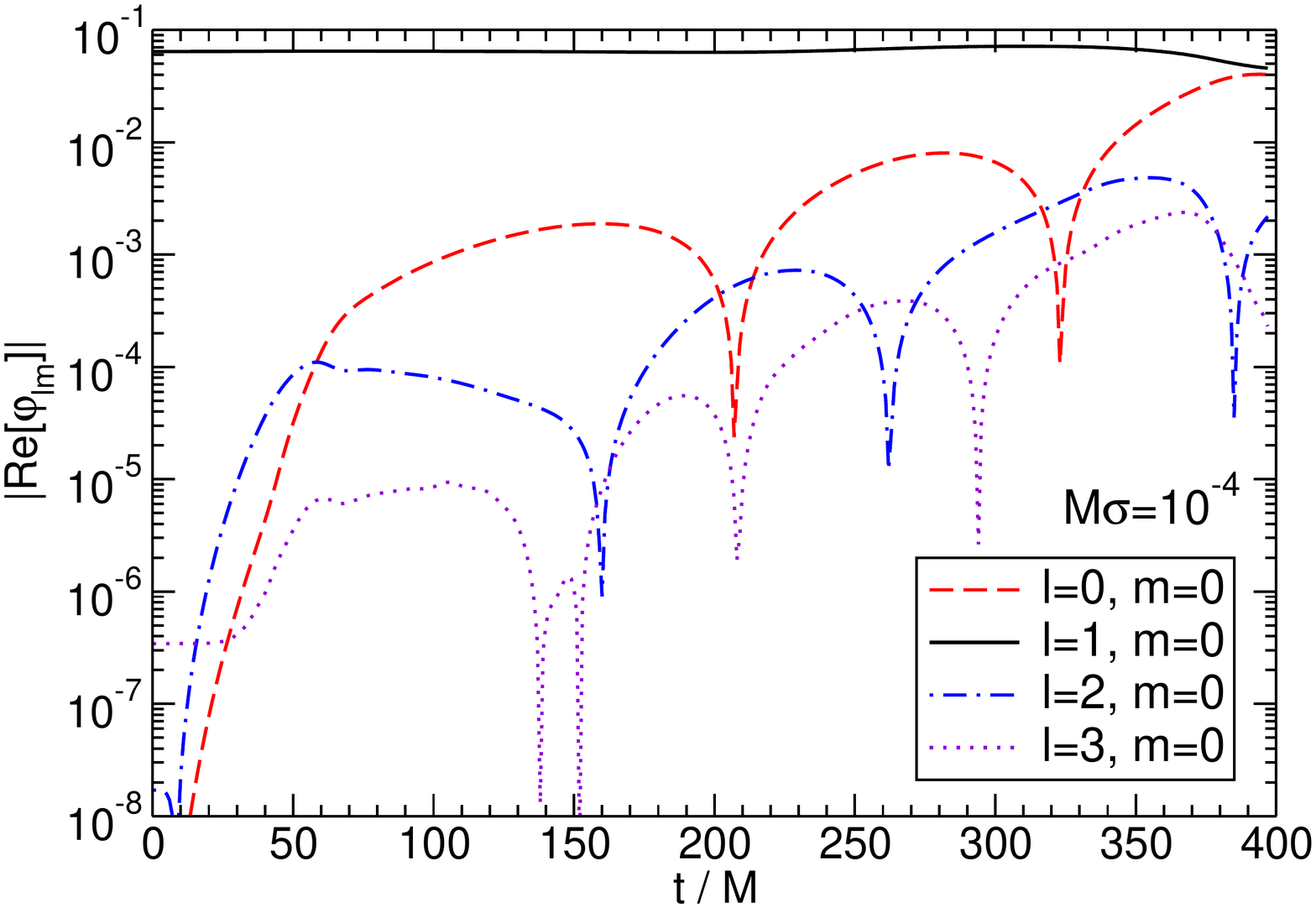}
\end{center}
\caption{Absolute value of the real part of the scalar multipoles
  $\left|{\rm Re}({\varphi}_{l0})\right|$ evaluated at the largest
  extraction radius $\tilde{r}=50M$ for different values of $l$ and
  two values of the scalar-field gradient, $M\sigma=10^{-5}$ and
  $M\sigma=10^{-4}$ (left and right panel, respectively).}
\label{fig:phi_sigma}
\end{figure*}

The spherical-harmonic expansion\footnote{This multipole expansion can
  be simply related to that introduced in
  Eq.~(\ref{eq:multipole_expansion}). For example, we have:
\be
{\dot{\vec{\cal D}}} =
\frac{r}{2} \sqrt{\frac{3}{\pi}}
 \Biggl[ {\varphi}_{10} \hat{x} 
- \frac{{\varphi}_{11} - {\varphi}_{1\,-1}}{\sqrt{2}} \hat{y}
-i \frac{{\varphi}_{11} + {\varphi}_{1\,-1}}{\sqrt{2}} \hat{z}\Biggr]_{1/r} \,,
\ee
where the subscript $1/r$ denotes that only the $1/r$ term should be
kept.} of the scalar field reads
\begin{equation}\label{phi_mode_expansion}
\varphi(t,r,\theta,\phi)=\sum_{lm}\varphi_{lm}(t,r)Y_{lm}(\theta,\phi)\,,
\end{equation}
where
\begin{equation}
\varphi_{lm}(t,r)=\int d\Omega\,\varphi(t,r,\theta,\phi)Y^*_{lm}(\theta,\phi)\,.
\end{equation}
Since the binary moves in the $yz$-plane, we use noncanonical rotated
coordinates
\begin{eqnarray}
\label{noncanonical}
x &=& r \cos \theta \,, \\
y &=& r \sin \theta \cos \phi \,, \nn \\
z &=& r \sin \theta \sin \phi \,. \nn
\end{eqnarray}

For small scalar-field gradients, the back-reaction on the metric is
very small and we expect to recover the stationary solution
$\varphi_{\rm ext}$ found in Section~\ref{sec:singlebh}, i.e., we
expect that after an initial transient
\begin{equation}
\varphi(t\rightarrow\infty,r,\theta,\phi)=\varphi_{\rm ext}(r,\theta)=
\varphi_0+2\pi\sigma(r-M)\cos\theta\,.
\end{equation}
where $\varphi_0$ is a constant. Because
$|Y_{10}|=\sqrt{3/(4\pi)}\cos\theta$,
the dominant nonvanishing component ${\varphi}_{lm}$ at late
times (up to quadratic corrections in the field) should be given by
\begin{align}
{\varphi}_{10}&=\sqrt{\frac{4\pi}{3}}2\pi\sigma(r-M)\,.
\label{an_prediction}
\end{align}
In this expression the areal radius $r$ is effectively time-dependent,
and it is computed dynamically during the evolution as described
above.

The numerical and analytical predictions (solid and dashed lines,
respectively) are compared in Figs.~\ref{fig:phi_sigma_allradii}
and \ref{fig:phi_sigma45}.
Let us focus first on Fig.~\ref{fig:phi_sigma_allradii}, which refers
to $M\sigma=10^{-5}$. In this case the numerically extracted dipole
mode asymptotes quickly to the analytical prediction. The inset shows
the percentage difference between the analytical and numerical value
of the scalar field at late times as a function of the extraction
radius: the agreement is remarkable at large radii but it gets
progressively worse as we get closer to the BH, most likely because of
gauge effects (we can exclude that the deviations are due to nonlinear
effects, because these would scale with $\sigma^2$, whereas the
disagreement seems to be independent of $\sigma$).
To summarize, the analytical and numerical predictions agree within a
few percent for gradients $M\sigma\leq 10^{-5}$.

The situation changes significantly for larger values of $M\sigma$. In
Fig.~\ref{fig:phi_sigma45} we overplot the numerical and analytical
components for $M\sigma=10^{-5}$ and for a larger gradient,
$M\sigma=10^{-4}$. As expected, when rescaled by their respective
gradients the dipolar components are essentially the same, and they
also converge to the value predicted by the analytical
solution. However for $M\sigma=10^{-4}$ this convergence can only be
seen at intermediate times, whereas at late times the mode develops an
instability: the figure shows that the roughly constant ``baseline''
value of the field seems to be superimposed to an exponentially
growing oscillation that develops on timescales $t\sim 10^2M$, both
when we evaluate the field numerically (continuous black line) and
when we use the areal radii to compute the analytical solution (dashed
black line).

The existence of this instability is confirmed by
Fig.~\ref{fig:phi_sigma}, where we look at higher multipoles of the
field on a logarithmic scale for $M\sigma=10^{-5}$ (left panel) and
for $M\sigma=10^{-4}$ (right panel). The right panel of this plot
shows that for the larger gradient $M\sigma = 10^{-4}$ an
exponentially growing instability with growth time $\sim 10^2M$ is
present also in the subdominant $l=0$, $l=2$ and $l=3$ multipoles. The
left panel illustrates that when $M\sigma = 10^{-5}$ the instability
(if it is present at all) develops on much longer timescales $t\gtrsim
10^3M$, so it does not affect our numerical simulations. Notice that
early-intermediate time results are physically consistent: for both
values of the gradient the scalar-field distribution is dominated by
the dipolar component, and it is in good agreement with analytical
predictions.

In summary, single-BH simulations in a scalar-field gradient show that
our numerical evolutions are stable and reliable as long as the
gradient is not too large. This conclusion is compatible with the
arguments presented in Section \ref{upperbound} above.

%%%%%%%%%%%%%%%%%%%%%%%%%%%%%%%%%%%%%%%%%%%%%%%%%%%%%%%%%%%%%%%%%%%%%%%%%%%%%%
\subsection{Binary black-hole evolutions: scalar and gravitational radiation}
\label{sec:numBBH}
%%%%%%%%%%%%%%%%%%%%%%%%%%%%%%%%%%%%%%%%%%%%%%%%%%%%%%%%%%%%%%%%%%%%%%%%%%%%%%
%

In this Section we discuss our numerical evolutions of BH binaries in
a scalar-field gradient.
In this initial study we focus on evolutions of nonspinning,
unequal-mass binaries with a gradient $M\sigma=2\times 10^{-7}$ and
mass ratio $q=3$, because unequal-mass binaries display two
interesting features which would be absent by construction in the
equal-mass case: center-of-mass recoil [see
  Eqs.~(\ref{eqB1})-(\ref{eqB3})] and monopole scalar radiation [see
  Eq.~(\ref{eqB4})].

\begin{figure*}[hbt]
\begin{center}
\includegraphics[width=0.48\textwidth]{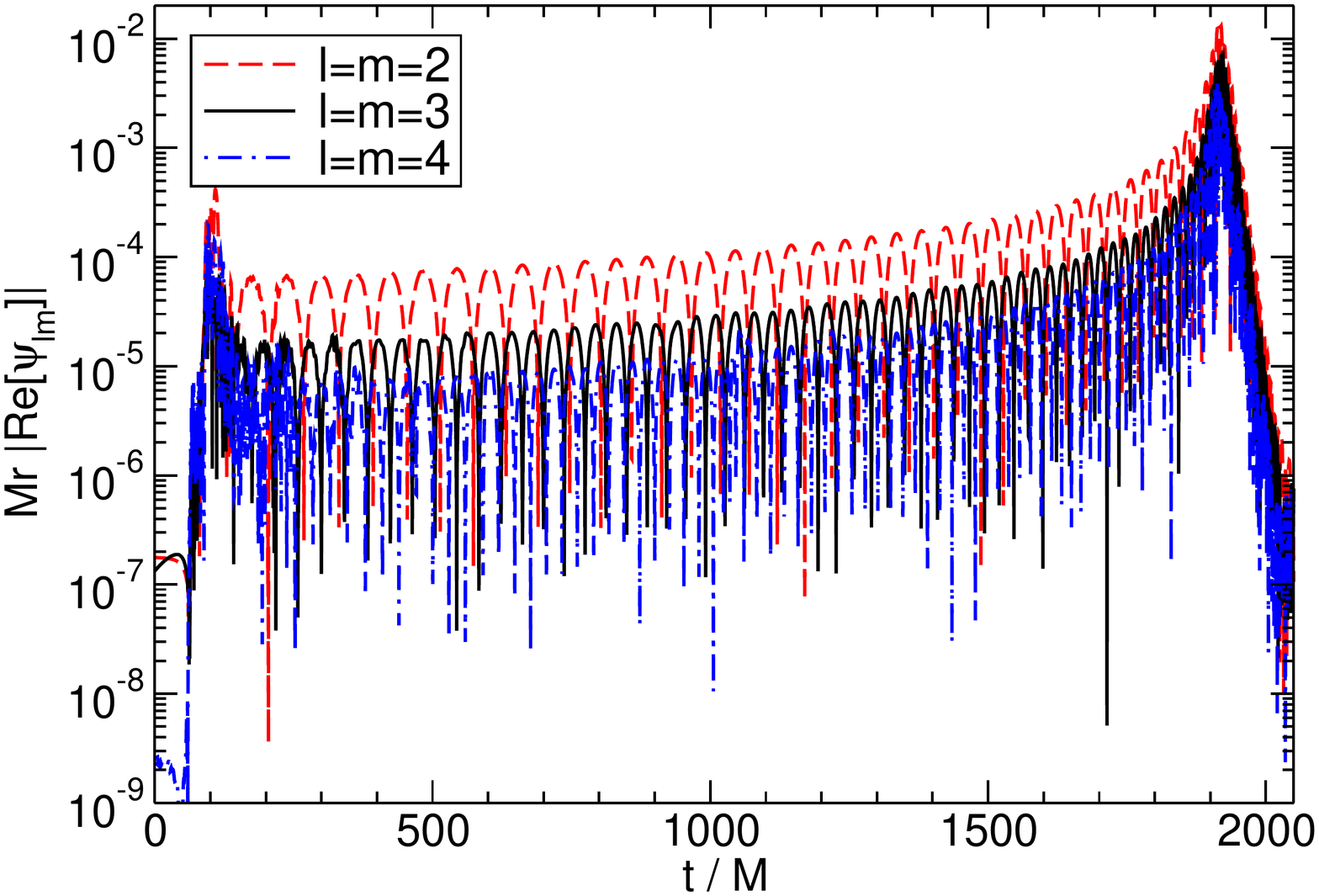}
\includegraphics[width=0.48\textwidth]{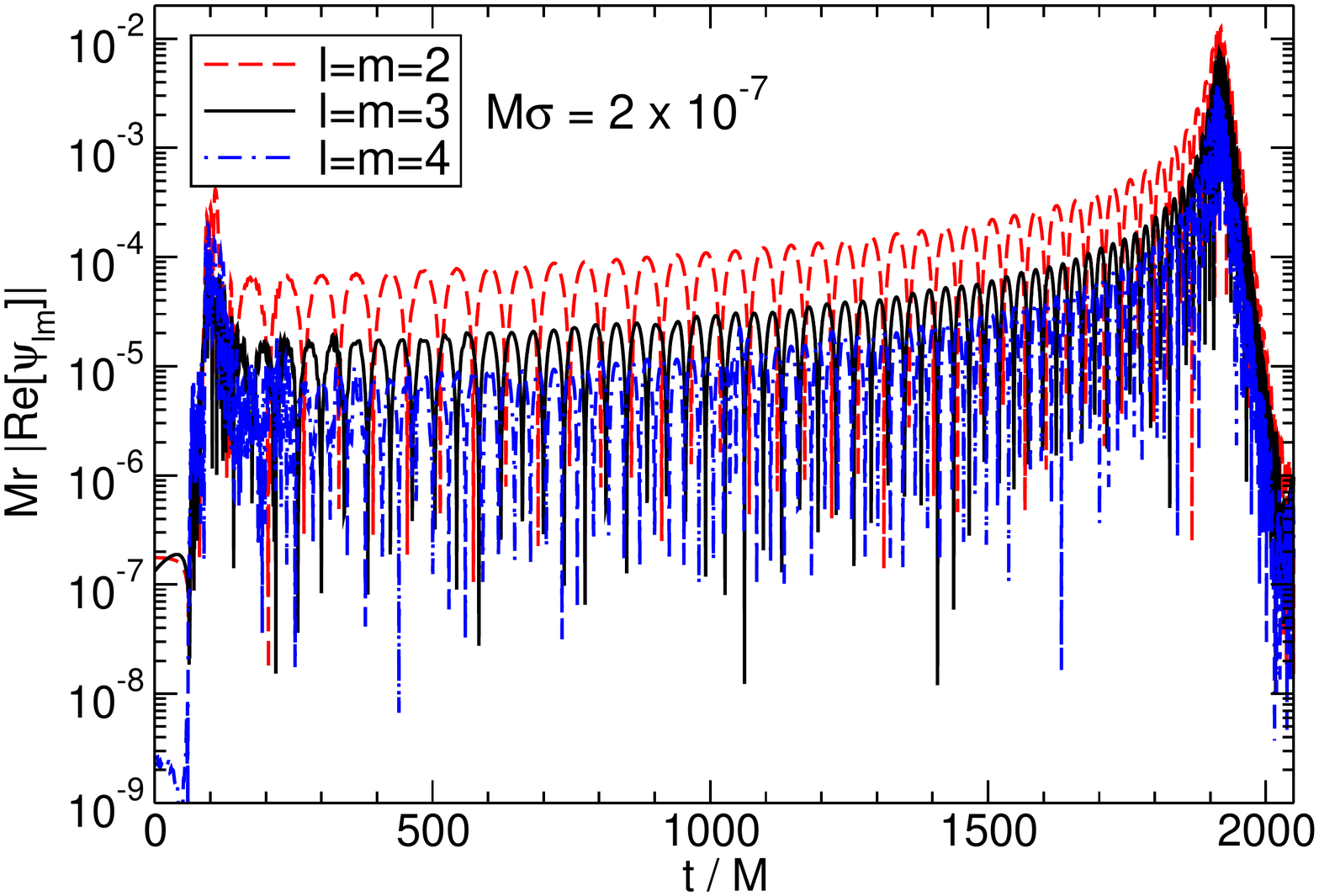}\\
\end{center}
\caption{Numerical results for a BH binary inspiralling in a scalar
  field gradient, with the orbital angular momentum perpendicular to
  the gradient. We show the spin-weighted spheroidal harmonic
  components of the Weyl scalar $\Psi_4$, $|{\rm Re}(\psi_{lm})|$,
  extracted at $r=56~M$ for $l=m$ (the imaginary parts are identical,
  modulo a phase shift). Left: $M \sigma=0$, right: $M \sigma=2\times
  10^{-7}$.}
\label{fig:psi4}
\end{figure*}

Gravitational waveforms, as characterized by the Newman-Penrose scalar
$\Psi_4$, are shown in Fig.~\ref{fig:psi4}. For such low values of the
scalar-field gradient, the impact on gravitational radiation emission
is hardly noticeable.

The emission of scalar radiation is much more interesting. The scalar
field acquires a nontrivial profile due to the dynamics of the
orbiting BH binary. The scalar radiation has a crucial dependence on
the binary setup, and more specifically on the angle between the
orbital angular momentum of the binary and the direction of the
scalar-field gradient. If this angle is zero, then effectively the
individual BHs do not traverse any field gradient, and the scalar
profile is expected to be trivial. Our numerical results confirm this
expectation: the output of these simulations is indistinguishable from
vacuum evolutions in pure general relativity
\cite{Berti:2007dg,Sperhake2010a}.

\begin{figure*}[hbt]
\begin{center}
\includegraphics[width=0.48\textwidth]{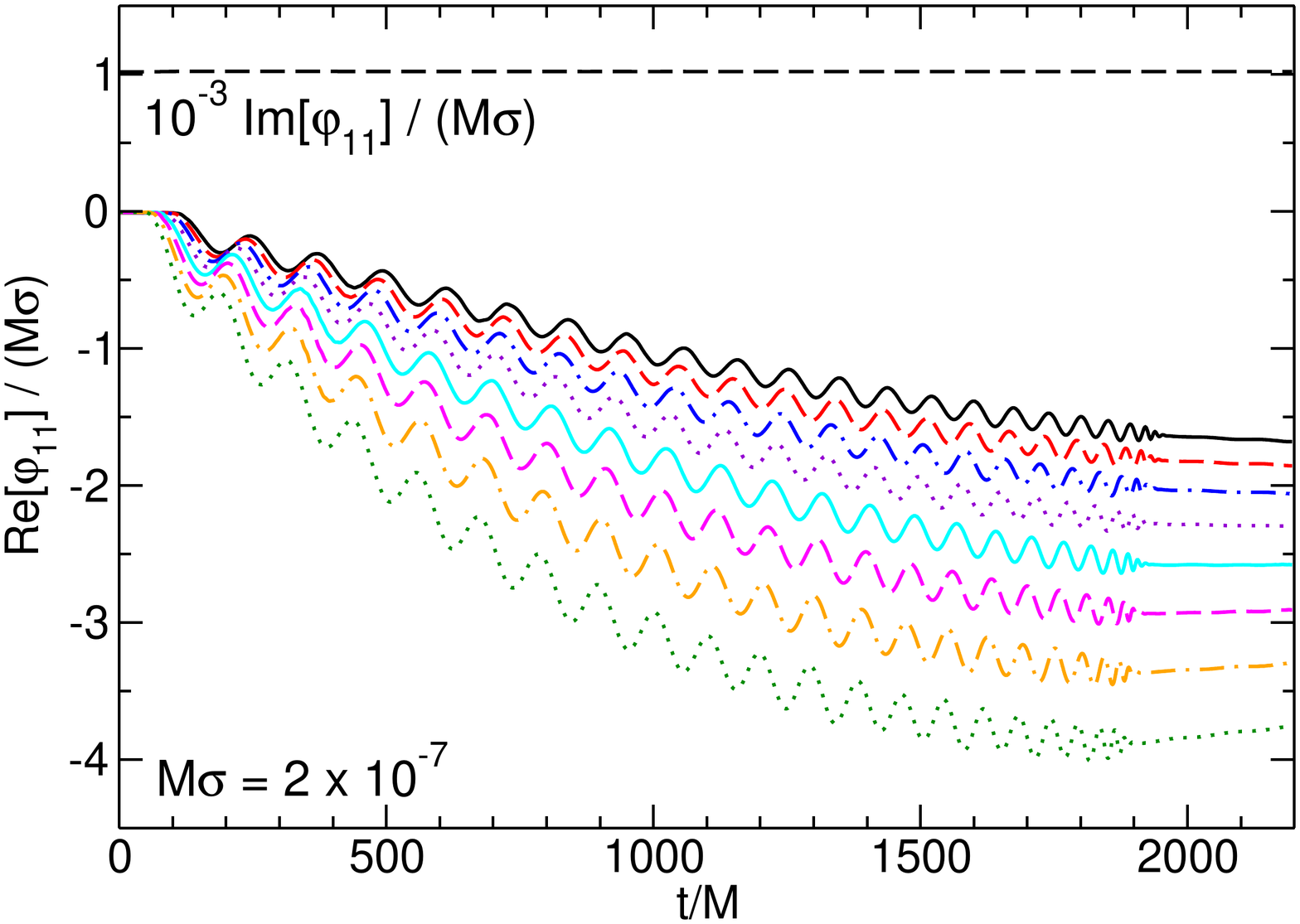}
\includegraphics[width=0.48\textwidth]{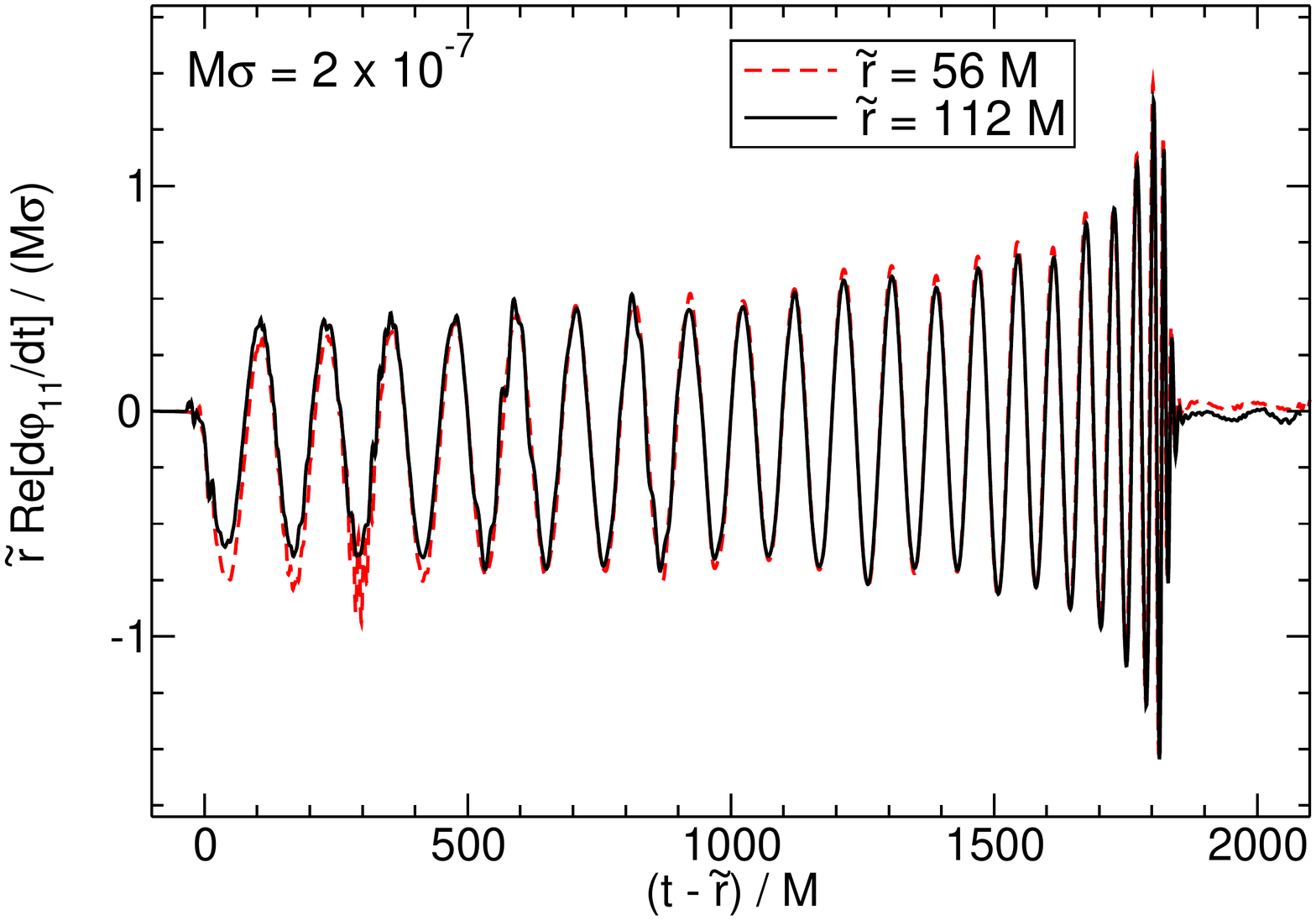}
\end{center}
\caption{Numerical results for a BH binary inspiralling in a scalar
  field gradient, with the orbital angular momentum perpendicular to
  the gradient.  Left: dependence of the various components of the
  scalar radiation ${\rm Re}(\varphi_{11})/(M\sigma)$ on the
  extraction radius (top to bottom: 112$M$ to 56$M$ in equidistant
  steps).  The dashed line corresponds instead to $10^{-3}{\rm
    Im}(\varphi_{11})/(M\sigma)$ at the largest extraction
  radius. Right: time-derivative of the scalar field at the largest
  and smallest extraction radii, rescaled by radius and shifted in
  time. Notice how the two waveforms show a clean and typical merger
  pattern, and that they overlap showing that the field scales to good
  approximation as $1/\tilde{r}$.}
\label{fig:philmall}
\end{figure*}

On the other hand, the induced scalar radiation should be maximized
when the orbital angular momentum is perpendicular to the field
gradient, so we now focus on this case. Our results are summarized in
Figs.~\ref{fig:psi4} and \ref {fig:philmall}.

Because the binary evolves on the background of a dipolar scalar-field
profile, this constant ``background'' value of the scalar shows up as
a large imaginary component\footnote{In both the single-BH solution
  (\ref{rotating_solution}) with $\gamma=\pi/2$ and in the numerical
  solution discussed here, the polar axis (in terms of which the polar
  angles, and then the harmonic decomposition, are defined) is
  orthogonal to the gradient. However the ordering of the axes is
  different in the two cases. This explains why the imaginary part
  ${\rm Im}(\varphi_{11})$ of the numerical solution corresponds to
  the real part ${\rm Re}(\varphi_{11})$ of the analytical solution.}
of the $l=|m|=1$ scalar-field modes, which is apparent in the left
panel of Fig.~\ref{fig:philmall} (in fact, we had to rescale the
imaginary component by a factor $10^{-3}$ in order to show this on
this plot).

The real part of $\varphi_{11}$ displays interesting dynamics (the
imaginary component also has similar dynamics, but this is partially
masked by the large background dipolar field, so the analysis of the
real part turns out to be numerically ``cleaner.'') At any extraction
radius ${\rm Re}(\varphi_{11})$ is initially zero, as the binary is
simply traversing a constant scalar field.  As the binary evolves, we
expect to see scalar radiation crossing the extraction surface and
producing a nonvanishing scalar profile. This is indeed observed in
Fig.~\ref{fig:philmall}, where we show that ${\rm Re}(\varphi_{11})$
has a ``wavy'' pattern at any fixed extraction radius. The scalar-wave
nature of this pattern is well illustrated by the right panel of
Fig.~\ref{fig:philmall}. There we take the time derivative of ${\rm
  Re}(\varphi_{11})$
at the largest and smallest extraction radii, scaling the amplitude of
the signal by the ratio of the extraction radii (as expected for a
wave scaling like $1/r$), and we observe that the signal is indeed
dipole scalar radiation emitted at twice the orbital frequency of the
binary, consistent with the predictions of Sec.~\ref{appsol}.
Furthermore, in Eq.~(\ref{Atheor}) of Appendix \ref{app:binaryscaleq}
we show that the amplitude of the $\varphi_{11}$ mode is consistent in
order of magnitude with analytical predictions. We also observe a
monopole component $\varphi_{00}$ whose amplitude is comparable to the
amplitude of $\varphi_{11}$, consistently with analytical predictions.

To understand the merger signal, when the two BHs collide and relax to
a final nearly stationary state, it is useful to remember that, in
vacuum, the merger of a BH binary with mass ratio $q=3$ produces a
Kerr BH with spin $a/M\sim 0.543$ \cite{Berti:2007dg}. Thus the lowest
ringdown frequencies are expected to be, from perturbative
calculations, $M\omega=0.351-0.0936i, 0.476-0.0849i$ for a $l=m=1$
scalar field, $l=m=2$ gravitational mode
\cite{Berti:2005ys,Berti:2009kk}. We find that $\Psi_4$ indeed rings
down with $M\omega \sim 0.48-0.081i$, in good agreement with
perturbative calculations. An analysis of $\dot{\varphi}_{11}$ yields
a ringdown frequency $M\omega \sim 0.36-0.070i$ (with errors $\lesssim
10\%$),
which is roughly consistent with perturbative calculations of scalar
($s=0$) perturbations of Kerr BHs \cite{Berti:2009kk}. Our simulations
also show that the ringdown phase, and indeed the entire scalar
signal, scales with $\sigma$. We conclude that our results indeed
represent linear effects, as opposed to nonlinear mode couplings.

Although not completely obvious, there is a small DC component in
Fig.~\ref{fig:philmall} (right panel), which we estimate to be
\begin{equation}
  \frac{|\dot{\varphi}^{DC}_{11}|}{|\dot{\varphi}^{\rm peak}_{11}|}\sim 0.2\,,
\end{equation}
at early times, where $|\dot{\varphi}^{\rm peak}_{11}|$ is the absolute
value of the waveform at a local peak (maximum or minimum).  This
numerical estimate can be compared to the analytical prediction,
Eqs.~(\ref{phiDCdip})-(\ref{phioscdip}). We start by estimating the
angular frequency $\dot{\chi}$ through the waveform frequency, and we
find $M\dot{\chi}\sim 0.025$.  Using Kepler's law, one can estimate
the orbital separation, and these two ingredients together with
relations (\ref{evolution_eqs}) allow us to estimate the relevant
ratio of the time derivatives of expressions
(\ref{phiDCdip})-(\ref{phioscdip}).  We find a ratio which is smaller
by almost one order of magnitude. This discrepancy can probably be
explained by numerical uncertainties and strong-field nonlinear
effects. 

It is apparent from Fig.~\ref{fig:philmall} (left panel) that the
${\rm Re}(\varphi_{11})$ modes display a ``drift'': after all the
dynamics has died away, the field does {\it not} return to zero. Our
data implies that at late times ${\rm Re}(\varphi_{11})\sim -2.7
\times 10^{-4}r^{ -1.24}$ for $M\sigma=2\times 10^{-7}$. The most
natural interpretation of this drift is related to the DC component,
Eq.~\eqref{app:dipole_eq}, which predicts a linear growth in time --
roughly the same dependence that can be seen in
Fig.~\ref{fig:philmall}. One should also bear in mind that the
analytical result is a slow-motion expansion, whereas the numerical
results cover only the highly dynamical, nonlinear merger signal; some
deviation from a perfectly linear dependence is therefore expected.

There are other possible contributions to such a drift. A second
possible contribution is due to the nonvanishing of ${\rm
  Im}(\varphi_{11})$ for the analytical solution
(\ref{rotating_solution}) with $\gamma=\pi/2$ (recall that in both the
single-BH solution (\ref{rotating_solution}) and in the numerical
solution discussed here, the polar axis is orthogonal to the gradient,
but the ordering of the axes is different, which explains why real and
imaginary parts of the modes are swapped). However, the $1/r$ piece of
that solution is orders of magnitude smaller than the amplitude of the
drift we observe numerically, and therefore unlikely to explain our
observations.

Another possible contribution comes from gravitational recoil. We are
simulating an unequal mass binary, which acquires a kick from the
origin of our coordinate axis as a result of the merger. The kick
introduces ``spurious'' multipolar components with respect to a frame
which is not comoving with the final BH.  However, an
order-of-magnitude estimate shows that this effect is unlikely to
explain the observed drift.
In order of magnitude, the kick contribution to $\varphi$ can be
estimated by looking at the terms in Eq.~(\ref{phiCMdip}), say
$\dot{\vec{{\cal D}}}_{\rm recoil}$, that are proportional to $v_{\rm
  CM}$: this yields
\be
\left|\f{\dot{\vec{{\cal D}}}_{\rm recoil}}{\tilde{r}}\right| \sim 
\f{M\sigma v_{\rm CM} v_{\rm orb}}{\tilde{r}/M}\,.
\ee
Here $M\sigma=2\times 10^{-7}$, the extraction radius is
$\tilde{r}/M\sim 10^2$, and $v_{\rm orb}$ is the orbital velocity.
The maximum recoil velocity from a nonspinning BH binary is of order
$v_{\rm CM} \sim 7 \cdot 10^{-4}$ \cite{Gonzalez2007}, so
$\left|\dot{\vec{{\cal D}}}_{\rm recoil}/\tilde{r}\right|\lesssim
10^{-12}$ even if $v_{\rm orb}$ approaches unity. This contribution to
the dipole radiation is way too small to account for a significant
portion of the drift seen in our simulations.

Finally, a frame-dragging effect
can also contribute with a nonzero drift for the scalar field. The
coalescing binary drags the inertial frames, inducing a local rotation
of the coordinate lines. This induces, near the binary, an apparent
rotation in the $y$-$z$ plane of the extracted scalar field, which
determines a nonvanishing real part of $\phi_{11}$. While the order of
magnitude of this effect is roughly consistent with our numerical
findings, the decay of the frame-dragging effect with extraction
radius is not consistent with our data. Therefore frame dragging is
not a dominant contribution to the observed drift.

While the DC component accounts for the order of magnitude of the
drift observed in our numerical simulations, most likely the observed
drift is due to a combination of the effects mentioned above, and
possibly others. In particular, by imposing constant-gradient boundary
conditions at finite distance from the binary during the evolution we
are effectively injecting energy into the system. This causes a growth
of the scalar field, which may contribute significantly to the
drift. This expectation should be confirmed by longer simulations
and/or by simulations where the ``plates'' generating the scalar
gradient are located further away from the binary. We hope to return
to this problem in future work.

%%%%%%%%%%%%%%%%%%%%%%%%%%%%%%%%%%%%%%%%%%%%%%%%%%%%%%%%%%%%%%%%%%%%%%%%%%%%%%%%
\section{Conclusions and outlook}
\label{sec:conclusions}
%%%%%%%%%%%%%%%%%%%%%%%%%%%%%%%%%%%%%%%%%%%%%%%%%%%%%%%%%%%%%%%%%%%%%%%%%%%%%%%%
We have investigated BH dynamics in external field profiles, by
considering the very simple example of a constant scalar-field
gradient. The broad features of our analysis should translate to
other, more general settings, at least as long as the external force
varies on length- or time-scales which are larger than the typical
binary separation. Our results are in agreement with linear or
slow-motion expansions, and show conclusively that black hole binaries
evolving in a nontrivial background produce interesting scalar and
gravitational-field dynamics.

As discussed in Appendix \ref{cosmo}, the scalar-field profiles
currently considered in scalar-field dark matter models correspond
roughly to $M \sigma\sim10^{-15}$ or less for a typical stellar-mass
BH with $M=10\,M_\odot$. Because scalar radiation is proportional to
the gradient, the experimental relevance of our setting for
gravitational radiation from BH binaries, as observable by Advanced
LIGO or similar instruments, seems negligible. However, it is
interesting that gravitational-wave observations may yield upper
bounds on scalar field gradients at all. Furthermore, our estimates in
Appendix \ref{cosmo} predict larger field gradients for supermassive
black holes with $M=10^9~M_{\odot}$; for such binary systems the
gradient could reach values as large as $M\sigma \sim 10^{-7}$, close
to the values studied in this work. Finally, strong field gradients
can be encountered in other -- albeit more speculative -- dark matter
configurations, such as supermassive boson stars
\cite{Macedo:2013qea}, so the possibility to come across this type of
signals should be seriously taken into account.  The remarkable
agreement we find between our numerical results and linearized
predictions indicates that the relatively small values of $M\sigma$
considered in our simulations fall into an effectively linear
regime. For the more speculative scenarios leading to $M\sigma$
significantly larger than the value $2\times 10^{-7}$ considered in
Sec.~\ref{sec:numBBH}, we therefore expect stronger numerical,
nonlinear effects to be present in the radiation.

Our analysis answers some questions and sparks many new ones: why
exactly do large gradients develop instabilities? Are these
instabilities of a physical or purely numerical nature? Even isolated
BHs moving in a scalar field should accrete: how can we understand the
details of this accretion process? Another interesting question
concerns spinning BHs. The original linearized analysis by Press
\cite{Press:1972} shows that Kerr BHs should in principle align their
rotation axis with the field gradient over long enough timescales. 
Numerical simulations of this alignment and of its nonlinear
development are an interesting (but numerically challenging) open
problem, which probably requires much longer simulations than those
presented in this work.

%%%%%%%%%%%%%%%%%%%%%%%%%%%%%%%%%%%%%%%%%%%%%%%%%%%%%%%%%%%%%%%%%%%%%%%%%%%%%%
\section*{Acknowledgments}
%%%%%%%%%%%%%%%%%%%%%%%%%%%%%%%%%%%%%%%%%%%%%%%%%%%%%%%%%%%%%%%%%%%%%%%%%%%%%%
E.B. and M.H. are supported by NSF CAREER Grant No. PHY-1055103.
V.C. acknowledges partial financial support provided under the
European Union's FP7 ERC Starting Grant ``The dynamics of black holes:
testing the limits of Einstein's theory'' grant agreement no.
DyBHo--256667, the NRHEP 295189 FP7-PEOPLE-2011-IRSES Grant, and
FCT-Portugal through projects PTDC/FIS/116625/2010,
CERN/FP/116341/2010 and CERN/FP/123593/2011.  Research at Perimeter
Institute is supported by the Government of Canada through Industry
Canada and by the Province of Ontario through the Ministry of Economic
Development and Innovation.
U.S. acknowledges support from
FP7-PEOPLE-2011-CIG Grant No.~293412 CBHEO,
the STFC GR Roller Grant No. ST/I002006/1,
the Ram\'on y Cajal Programme and Grant FIS2011-30145-C03-03 of
the Ministry of Education and Science of Spain.
Computations were performed on the ``Baltasar Sete-Sois'' cluster at
IST, the cane cluster in Poland through PRACE DECI-7 ``Black hole
dynamics in metric theories of gravity'', on Altamira in Cantabria
through BSC grant AECT-2012-3-0012, on Caesaraugusta in Zaragoza
through BSC grants AECT-2012-2-0014 and AECT-2012-3-0011, XSEDE
clusters SDSC Trestles and NICS Kraken through NSF
Grant~No.~PHY-090003, Finis Terrae through Grant CESGA-ICTS-234 and
the COSMOS supercomputer, part of the DiRAC HPC Facility which is
funded by STFC and BIS. We thank Andrey Kaliazin for computational
support and technical advice, and Ivan Stefanov for drawing our
attention to references
\cite{Stefanov:2007qw,Stefanov:2007eq,Stefanov:2007bn,Stefanov:2009qd,Stefanov:2009zza,Doneva:2010ke}.

%%%%%%%%%%%%%%%%%%%%%%%%%%%%%%%%%%%%%%%%%%%%%%%%%%%%%%%%%%%%%%%%%%%%%%%%%%%%%%

\appendix
%%%%%%%%%%%%%%%%%%%%%%%%%%%%%%%%%%%%%%%%%%%%%%%%%%%%%%%%%%%%%%%%%%%%%%%%%%%%%%
\section{Boosted black-hole background}
\label{app:boosted}
%%%%%%%%%%%%%%%%%%%%%%%%%%%%%%%%%%%%%%%%%%%%%%%%%%%%%%%%%%%%%%%%%%%%%%%%%%%%%%
In this Appendix we shall consider the solution of the Klein-Gordon
equation in a boosted BH background, i.e., in the presence of a
Schwarzschild BH moving with constant velocity $v$ in the direction
orthogonal to the charged planes. We shall show that this solution
does not emit scalar radiation: only accelerated BHs moving in a
uniform scalar-field gradient can emit scalar radiation.

To this aim, we shall first consider the (regular) solution describing
a scalar field on a Schwarzschild background, generated by an infinite
plane moving with velocity $v$ along the direction $z$ orthogonal to
the plane. Then, we shall boost back this solution, to obtain a moving
BH and a scalar field generated by a fixed plane.

The expression
\be
\varphi=2\pi\sigma \gamma
\left[(r-M)\cos\theta+2Mv\left(\frac{V-r}{2M}-\log\frac{r}{2M}\right)\right]
\label{movingphi}
\ee
describes a solution, regular at the horizon, of the Klein-Gordon
equation (\ref{scfeq}) on a Schwarzschild background. Here
$r_*=r+2M\log(r/2M-1)$ is the tortoise coordinate, $V=t+r_*$ is the
standard advanced time coordinate, $v$ is a velocity parameter and
$\gamma=(1-v^2)^{-1/2}$.  For $v=0$, the previous expression reduces
to the static solution (\ref{defphiext}). For finite $v$ and at
asymptotically large distances it reduces to
\be
\varphi \sim 2\pi\sigma \gamma (z+vt)\,,
\ee
which is related to the asymptotic solution (\ref{defphiext}) by a
simple boost.  Therefore, Eq.~(\ref{movingphi}) describes a scalar
field (on a Schwarzschild background) generated by an infinite plane
moving with velocity $-v$ along $z$.  We shall now boost
Eq.~(\ref{movingphi}) and the Schwarzschild metric, in order to find a
solution of the field equations describing a BH moving with velocity
$v$ in a field generated by an infinite charged plane at rest. The
boost is more easily performed in isotropic coordinates
$(t,\tr,\theta,\phi)$, in which
\begin{align}
\varphi&=2\pi\sigma \gamma\left[z\left(1+
\frac{M^2}{4\tr^2}\right)+tv\right.\nn\\
&\left.-2Mv \log{\frac{4
+M/\tr+4\tr/M}{-4+M/\tr+4\tr/M}}\right]\,.
\end{align}
If we apply a Lorentz boost along the $z$-direction
\begin{align}
\bar{t}&=\gamma(t+vz)\nn\\
\bar{z}&=\gamma (z+vt)\nn\\
\bar{x}&=x\nn\\
\bar{y}&=y\,,
\end{align}
we get
\begin{align}
\varphi&=2\pi\sigma\left[\bar{z}+\frac{\gamma^2 M^2}{4\tr^2}(\bar z-v\bar
  t)\right.\nn\\
&\left.-2Mv\gamma \log{\frac{4+M/\tr+4\tr/M}{-4+M/\tr+4\tr/M}}\right]\,.
\end{align}
Note that the isotropic radial coordinate reads
$\tr^2=\bar{x}^2+\bar{y}^2+\gamma^{2}(\bar{z}-v\bar{t})^2$.

Along the $\bar{z}-$axis, this expression has the form
\be
\varphi\sim 2\pi\sigma \left[\bar{z}
+\frac{M^2(1-16v)}{4\bar{z}}
+\frac{M^2(1-16v)v\bar{t}}{4\bar{z}^2}\right]+
{\cal O}(\bar{z}^{-3})\,.
\ee
Introducing polar coordinates in the boosted frame
\begin{align}
\bar x&=\bar r\sin\bar\theta\cos\bar\phi\nonumber\,,\\
\bar y&=\bar r\sin\bar\theta\sin\bar\phi\nonumber\,,\\
\bar z&=\bar r\cos\bar\theta\,,
\end{align}
we have (since $\gamma^2-1=v^2\gamma^2$)
\begin{equation}
\tr=\sqrt{{\bar r}^2(1+v^2\gamma^2\cos\bar\theta)-2v\gamma^2\bar t
\bar r\cos\bar\theta+\gamma^2v^2\bar t^2}\,.\label{trbr}
\end{equation}
If we assume small boosts $v\ll1$, so we can neglect terms $O(v^2)$,
$\gamma\simeq1$.  We can also assume that $v\bar t\ll\bar r$, i.e.,
that $v\bar t/\bar r\ll 1$, even though this quantity can be larger
than $v^2$. Expanding Eq.~(\ref{trbr}) in these two dimensionless
quantities, up to first order in $v^2$ and second order in $v\bar
t/\bar r\ll 1$), we find
\begin{equation}
\tr\simeq\bar r\left[1-\left(\frac{v\bar t}{\bar r}\right)\cos\bar\theta
+\frac{1}{2}\left(\frac{v\bar t}{\bar r}\right)^2\right]\,.
\end{equation}
Therefore
\begin{align}
\varphi&=2\pi\sigma\left[\bar r\cos\bar\theta+\frac{M^2}{4\bar r}
\left(\cos\bar\theta+2\left(\frac{v\bar t}{\bar r}\right)\cos^2\bar\theta\right)
\right.\nn\\
&\left.-\frac{4M^2v}{\bar r}\left(1+\left(\frac{v\bar t}{\bar r}\right)
\cos\bar\theta\right)\right]+O\left(\frac{1}{{\bar r}^3}\right)\nonumber\\
&=2\pi\sigma\left[\bar r\cos\bar\theta+\frac{M^2}{4\bar r}
(\cos\bar\theta-16v)\right.\nn\\
&\left.+\frac{M^2 v\bar t}{2{\bar r}^2}\cos\bar\theta
(\cos\bar\theta-8v)\right]\,.
\end{align}
As discussed in \cite{Horbatsch:2011ye}, we find that the scalar charge is the
(spherically symmetric component of the) coefficient of $1/{\bar r}$
in this expansion, divided by $2M$, thus
\begin{equation}
Q=-4\pi\sigma Mv\,.
\end{equation}
Furthermore, the scalar-field multipoles 
are
\begin{align}
{\cal M}&=8\pi\sigma M^2v\,,\\
{{\dot{\vec{\cal D}}}}&=-\frac{\pi\sigma M^2}{2}\hat z\,.
\end{align}
All of these quantities are constant in time, therefore there is no
emitted scalar radiation.
We can conclude that if a BH moves with constant velocity in a
scalar-field gradient, there is no scalar emission; in order to get
scalar radiation the BH should have a nonvanishing acceleration, as
shown analytically in Section \ref{anbin} and numerically in Section
\ref{sec:numBBH}.

%%%%%%%%%%%%%%%%%%%%%%%%%%%%%%%%%%%%%%%%%%%%%%%%%%%%%%%%%%%%%%%%%%%%%%%%%%%%%%
\section{Approximate solution for monopole and dipole radiation from a quasi-circular binary in a scalar gradient}
\label{app:binaryscaleq}
%%%%%%%%%%%%%%%%%%%%%%%%%%%%%%%%%%%%%%%%%%%%%%%%%%%%%%%%%%%%%%%%%%%%%%%%%%%%%%
This Appendix completes the approximate solution discussed in Section
\ref{appsol}, which describes the scalar field generated by a binary
BH in quasicircular orbit in a scalar-field gradient.  The
center-of-mass acceleration is:
\begin{eqnarray}
\label{eqB1}
\vec{a}_{\rm CM} &=& \frac{4a^{(0)}\dot{\chi}^{(0)}}{5 \tau_{\rm q}} \frac{(q-1)}{(q+1)} 
\left[ \hat{z} \cos \chi^{(0)} - \hat{y} \sin \chi^{(0)} \right] \,,
\\
\label{eqB2}
\vec{v}_{\rm CM} &=& \frac{4a^{(0)}}{5 \tau_{\rm q}}\frac{(q-1)}{(q+1)} 
\left[ \hat{y}(\cos \chi^{(0)}-1) + \hat{z} \sin \chi^{(0)} \right] 
\nonumber
\\
&&+ \vec{V}_{0} \,,
\\
\label{eqB3}
\vec{z}_{\rm CM} &=& \frac{4a^{(0)}}{5 \tau_{\rm q} \dot{\chi}^{(0)}} \frac{(q-1)}{(q+1)} 
\left[ \hat{y} \sin \chi^{(0)} - \hat{z} (\cos \chi^{(0)}-1) \right] 
\nonumber
\\
&&
+ \vec{V}_{0}(t-t_{0}) \,,
\end{eqnarray}
where $q=M_{1}/M_{2}$ is the mass ratio, $\chi^{(0)}(t) =
\dot{\chi}^{(0)}t + \psi_{0}$ is the zeroth-order orbital phase, which
vanishes at $t=t_{0}$, and $\vec{V}_{0}=\vec{v}_{\rm CM}(t=t_{0})$ is
the initial velocity of the center of mass relative to the scalar
gradient.  Without loss of generality, the choice $\vec{z}_{\rm
  CM}(t=t_{0}) = 0$ has been made.  Note that in the equal-mass limit
($q \to 1$), the center-of-mass recoil vanishes, on account of
symmetry.

Given the scalar charges
\begin{eqnarray} 
Q_{1} &=& 
\frac{8 \pi \sigma M_{1}^{2}}{M} \biggl(  M\vec{v}_{\rm CM} \cdot \hat{z} 
+ M_{2} \left[ \dot{a} \sin \chi + a \dot{\chi} \cos \chi \right]\biggr) \,,
\nn
\\
Q_{2} &=&  
\frac{8 \pi \sigma M_{2}^{2}}{M} \biggl( M\vec{v}_{\rm CM} \cdot \hat{z}
- M_{1} \left[ \dot{a} \sin \chi + a \dot{\chi} \cos \chi \right]\biggr)\,,
\nn
\end{eqnarray}
the monopole component of the scalar field is:
\begin{widetext}
\begin{eqnarray}
\label{eqB4}
{\cal M}
 = Q_{1} + Q_{2}  = 
%\frac{8 \pi G \sigma}{M} 
\frac{8 \pi \sigma}{M} 
\biggl( M(M_{1}^{2} + M_{2}^{2})(\vec{v}_{\rm CM} \cdot \hat{z})
+ M_{1}M_{2}(M_{1}-M_{2})[\dot{a} \sin \chi + a \dot{\chi} \cos \chi] \biggr) \,,
\end{eqnarray}
while the dipole component reads
\be
\label{app:dipole_eq}
\vec{\cal D} = Q_{1}\vec{z}_{1}(t) + Q_{2} \vec{z}_{2}(t)
= \vec{\cal D}_{\rm CM} + 
\vec{\cal D}_{\rm rel,\, DC} 
+ \vec{\cal D}_{\rm rel,\, osc} \,,
\ee
with
\beq
\label{app:dipole_eq1}
\vec{\cal D}_{\rm CM} &=&
\frac{8 \pi \sigma M_{1}M_{2}(M_{1}-M_{2})a}{M} 
\biggl(
\vec{z}_{\rm CM} \dot{\chi} \cos \chi 
+ (\vec{v}_{\rm CM} \cdot \hat{z})( \hat{y} \cos \chi + \hat{z} \sin \chi )
\biggr) \,,
\\
\label{app:dipole_eq2}
\vec{\cal D}_{\rm rel , \, DC} &=& 
\frac{8 \pi \sigma M_{1}^{2}M_{2}^{2}a}{M^{2}} 
\biggl(
\dot{a} \hat{z}  + a \dot{\chi} \hat{y}
\biggr) \,,
\\
\label{app:dipole_eq3}
\vec{\cal D}_{\rm rel , \, osc} &=& 
\frac{8 \pi \sigma M_{1}^{2}M_{2}^{2}a}{M^{2}} 
\biggl(
\dot{a} [ \hat{y} \sin(2\chi)  - \hat{z} \cos(2\chi) ]
+ a \dot{\chi} [ \hat{y} \cos ( 2 \chi) + \hat{z} \sin (2 \chi) ]
\biggr) \,.
\eeq
From the previous expression we find
\be
\label{app:dipoledot_eq}
\dot{\vec{\cal D}} = 
\dot{\vec{\cal D}}_{\rm CM} + 
\dot{\vec{\cal D}}_{\rm rel,\, DC} + 
\dot{\vec{\cal D}}_{\rm rel,\, osc} \,,
\ee
with
\begin{eqnarray}
\label{phiCMdip}
\dot{\vec{\cal D}}_{\rm CM} &=&
\frac{8 \pi \sigma M_{1}M_{2}(M_{1}-M_{2})a}{M}
\biggl( 
\dot{\chi}(\vec{v}_{\rm CM} \cos \chi -  \vec{z}_{\rm CM} \dot{\chi} \sin \chi)
+ (\vec{a}_{\rm CM} \cdot \hat{z}) (\hat{y}\cos \chi + \hat{z} \sin \chi)
\\
&&
\phantom{\frac{8 \pi \sigma M_{1}M_{2}(M_{1}-M_{2})a}{M^{2}}
\biggl(}
+  (\vec{v}_{\rm CM} \cdot \hat{z}) \dot{\chi} (\hat{z} \cos \chi - \hat{y} \sin \chi)
\biggr) \,,
\nn
\\
\label{phiDCdip}
\dot{\vec{\cal D}}_{\rm rel, \, DC} &=&
\frac{8 \pi \sigma M_{1}^{2} M_{2}^{2}a}{M^{2}}
\biggl(
\ddot{a} \hat{z} + (2\dot{a}\dot{\chi} + a \ddot{\chi})\hat{y}
\biggr) \,,
\\
\label{phioscdip}
\dot{\vec{\cal D}}_{\rm rel, \, osc} &=&
\frac{8 \pi \sigma M_{1}^{2} M_{2}^{2} a}{M^{2}} 
\biggl(
(\ddot{a} - 2a\dot{\chi}^{2})[\hat{y}\sin ( 2\chi) - \hat{z} \cos (2 \chi) ]
+ (4 \dot{a} \dot{\chi} + a \ddot{\chi})[\hat{y} \cos ( 2 \chi) + \hat{z} \sin ( 2 \chi) ]
\biggr) \,,
\end{eqnarray}
\end{widetext}
where $M=M_{1} + M_{2}$ is the total mass, $\vec{v}_{\rm CM}$ and
$\vec{a}_{\rm CM}$ are the velocity and acceleration of the center of
mass, respectively, and terms of order higher than $1/c^{5}$ in the
dipole moment have been dropped.
If we neglect radiative effects, and express the dipole waveform using the $\ell, m$ multipole components introduced in (\ref{phi_mode_expansion}), we find
\begin{eqnarray}
\varphi_{10} &=& 0 \,,
\\
\varphi_{1\, \pm 1} &=& i A e^{\mp 2 i \chi}  + {\rm gradient\ term}\,,
\end{eqnarray}
where the amplitude of oscillation is given by
\begin{equation}
\label{Atheor}
A = \sqrt{\frac{512 \pi^{3}}{3}} (M \sigma) \frac{M}{r} \nu^{2} (M \dot\chi)^{2/3} \,,
\end{equation}
and $\nu=M_1M_2/M^2$ is the symmetric mass ratio. For our simulation
with $q=3$ and $M\sigma = 2\times 10^{-7}$ we find that $M \dot \chi
\sim 2\times 10^{-2}$, and therefore the theoretical prediction for
the dipole amplitude is $A \sim 4\times 10^{-10}$. This is in
order-of-magnitude agreement with the observed dipole radiation in our
numerical simulations.

%%%%%%%%%%%%%%%%%%%%%%%%%%%%%%%%%%%%%%%%%%%%%%%%%%%%%%%%%%%%%%
\section{3+1 Evolution equations}
\label{sec:evolution_equations}
%%%%%%%%%%%%%%%%%%%%%%%%%%%%%%%%%%%%%%%%%%%%%%%%%%%%%%%%%%%%%%

In terms of the BSSN variables defined in Eq.~(\ref{eq:BSSNvars}),
the scalar field $\varphi$ and the scalar curvature $K_{\varphi}$
defined in Eq.~(\ref{eq:Kphi}), the BSSN evolution equations
are given by
\begin{widetext}
\begin{eqnarray}
  \partial_t \tilde{\gamma}_{ij} &=& \beta^m \partial_m \tilde{\gamma}_{ij}
        + 2\tilde{\gamma}_{m(i} \partial_{j)} \beta^m - \frac{2}{3}
        \tilde{\gamma}_{ij} \partial_m \beta^m -2\alpha \tilde{A}_{ij}\,,
        \label{eq:gammat} \\
  \partial_t \chi &=& \beta^m \partial_m \chi + \frac{2}{3} \chi
      (\alpha K - \partial_m \beta^m)\,,
      \label{eq: chi} \\
  \partial_t \tilde{A}_{ij} &=& \beta^m \partial_m \tilde{A}_{ij}
        + 2\tilde{A}_{m(i} \partial_{j)} \beta^m - \frac{2}{3} \tilde{A}_{ij}
        \partial_m \beta^m + \chi \left( \alpha R_{ij}
        - D_i D_j \alpha\right)^{\rm TF} + \alpha \left( K\,\tilde{A}_{ij}
        - 2 \tilde{A}_i{}^m \tilde{A}_{mj} \right)
        \nonumber \\
     && -\alpha \chi \left( \partial_i \varphi \, \partial_j \varphi
        - \frac{1}{3} \tilde{\gamma}_{ij} \tilde{\gamma}^{mn}
        \partial_m \varphi \, \partial_n \varphi \right) \,, \\
  \partial_t K &=& \beta^m \partial_m K - D^m D_m \alpha + \alpha \left(
        \tilde{A}^{mn} \tilde{A}_{mn} + \frac{1}{3} K^2 \right)
        + 4\alpha K_{\varphi}\,,
        \label{eq:tracekt} \\
  \partial_t \tilde{\Gamma}^i &=& \beta^m \partial_m \tilde{\Gamma}^i
        - \tilde{\Gamma}^m \partial_m \beta^i
        + \frac{2}{3} \tilde{\Gamma}^i \partial_m \beta^m
        + 2 \alpha \tilde{\Gamma}^i_{mn} \tilde{A}^{mn}
        + \frac{1}{3} \tilde{\gamma}^{im}\partial_m \partial_n \beta^n
        + \tilde{\gamma}^{mn} \partial_m \partial_n \beta^i \nonumber \\
     && - \frac{4}{3} \alpha \tilde{\gamma}^{im} \partial_m K
        - \tilde{A}^{im} \left( 3 \alpha \frac{\partial_m \chi}{\chi}
          + 2\partial_m \alpha \right)
        - \frac{2}{3} \left(\tilde{\Gamma}^i
          -\tilde{\gamma}^{mn}\tilde{\Gamma}^i_{mn} \right)
          \partial_k \beta^k
        - 4 K_{\varphi} \tilde{\gamma}^{im}
        \partial_m \varphi\,.
        \label{eq:Gammat}
\end{eqnarray}
\end{widetext}

Likewise, we write the Hamiltonian and momentum constraints in terms
of the BSSN variables as
\begin{eqnarray}
  && ^{(3)}R+\frac{2}{3}K^2-{\tilde\gamma}^{mn}
     {\tilde\gamma}^{kl}{\tilde A}_{mk}
     {\tilde A}_{nl} = 4K_\varphi^2+\partial_i\varphi\partial^i\varphi\,,
     \label{eq:Ham} \\
  && \tilde{D}_m \tilde{A}^m{}_i - \frac{2}{3} \partial_i K
     -\frac{3}{2\chi} \tilde{A}^m{}_i \partial_m \chi
     = 2K_\varphi\partial_i\varphi\,.
     \label{eq:mom}
\end{eqnarray}

The lapse function and shift vector are evolved as in the case of
vacuum general relativistic BH simulations (cf.~\cite{vanMeter2006})
according to
\begin{eqnarray}
  \partial_t \alpha &=& \beta^m \partial_m \alpha -2\alpha K\,,
      \label{eq:alphat} \\
  \partial_t \beta^i &=& \beta^m \partial_m \beta^i
      + \frac{3}{4} \tilde{\Gamma}^i - \eta \beta^i\,.
      \label{eq:betat}
\end{eqnarray}
Following \cite{Schnetter:2010cz}, we use a position-dependent
parameter $\eta$; specifically, we set
\begin{equation}
  \eta = \eta_0 \frac{R^2}{r^2+R^2}
         \frac{|\vec{r}_{\rm 1}| + |\vec{r}_{\rm 2}|}
              {2(M_{\rm 1} |\vec{r}_{\rm 1}| + M_{\rm 2} |\vec{r}_{\rm 2}|)}\,,
         \label{eq:eta_beta}
\end{equation}
where $r$ is the coordinate distance from the origin, $\vec{r}_{\rm
  1,2}$ are the position vectors from either hole and $M_{\rm 1,2}$
are the BH masses. Lapse and shift are initialized as $\alpha =
\sqrt{\chi}$ and $\beta^i=0$, respectively.

%%%%%%%%%%%%%%%%%%%%%%%%%%%%%%%%%%%%%%%%%%%%%%%%%%%%%%%%%%%%%%%%%%%%%%%%%%%%%%
\section{Order of magnitude of the scalar gradient in a cosmological scenario}
\label{cosmo}
%%%%%%%%%%%%%%%%%%%%%%%%%%%%%%%%%%%%%%%%%%%%%%%%%%%%%%%%%%%%%%%%%%%%%%%%%%%%%%

Scalar fields on galactic scales have been considered by many authors
as a possible explanation for the rotation curves in galaxies and as
alternatives to cold dark matter
\cite{PhysRevD.50.3650,Hu:2000ke,Matos:1998vk,Schunck:1998nq,Sahni:1999qe,Matos:2000ki,Arbey:2001qi,Arbey:2003sj,Boehmer:2007um}
(see also \cite{Alcubierre:2001ea}). The aim of this Appendix is to estimate the
typical magnitudes of the scalar-field gradients predicted by these models.

%The rotational velocity near galactic centers $v_g\sim 200$~km/s is
%approximately constant \cite{Salucci:2002jg}. The mass $M(r)$
%contained within an orbit of radius $r$ is roughly given by
%\begin{equation}
%v_g^2\sim
%\frac{M(r)}{r}\,.
%\end{equation}
%Then, in order of magnitude, the density is 
%\begin{equation}
%\rho\sim
%\frac{v^2_g}{r^2}\,,
%\end{equation}
%on scales of the order $r\sim 10$ kpc $=3\times10^{17}$ km (see
%e.g. \cite{Boehmer:2007um}).  This density is provided by dark matter,
%or, in the scalar-field scenario, by the scalar field.

Sadeghian {\it et al.} \cite{Sadeghian:2013laa} recently studied the
distribution of dark matter around massive BHs in full general
relativity using a Hernquist profile \cite{Hernquist:1990be},
which is a good description of isolated dark matter halos
\cite{Visbal:2012ta}.
%, instead of the Navarro, Frank and White \cite{Navarro:1996gj}
%profile.
%
According to \cite{Sadeghian:2013laa}, a typical density for the dark
matter halo is $\rho\sim 10^{10}$~Gev/cm$^3$.
One should be very cautious in comparing our stationary, free scalar
field configuration with those suggested by cosmological
models. Indeed, in many of the works cited above the scalar field is
rapidly oscillating, and the mass term and the potential always play a
role. In order to estimate the order of magnitude of the scalar-field
gradient we can simply note that, neglecting the contribution of the
potential $V(\phi)$ and restoring physical units, the mass-energy
density is of the order (see e.g. \cite{Schunck:1998nq})
\begin{equation}
G\rho\sim|\phi_{,t}|^2+c^2|\phi_{,r}|^2\,,
%\rho\sim|\phi_{,t}|^2+c^2|\phi_{,r}|^2\,;
\end{equation}
therefore the gradient $\sigma=\phi_{,r}$ is at most
\begin{equation}
\sigma\sim
\sqrt{\frac{G\rho}{c^2}}
%\sim\frac{v_g}{r\,c}\sim 10^{-21}\,{\rm km}^{-1}\sim \frac{10^{-20}}{10 M_{\odot}}\,.
\sim 10^{-16}\,{\rm km}^{-1}\sim \frac{10^{-15}}{10 M_{\odot}}\,.
\label{estsigma}
\end{equation}
In our simulations we set the BH mass $M=1$. For a stellar-mass BH
($M=10\,M_\odot$) moving near the galactic center a typical
scalar-field gradient is therefore $M\sigma\sim 10^{-15}$; for a
supermassive BH ($M=10^9\,M_\odot$), a typical gradient would be
$M\sigma\sim 10^{-7}$, the same order of magnitude studied in this
paper. This should be considered as a rough upper limit: in scenarios
in which the scalar field is rapidly oscillating the kinetic term
should contribute to the energy density more than the gradient.

%\bibliographystyle{myutphys} %Needed by Leonardo
%\bibliography{ema}


\end{document}